\RequirePackage[l2tabu, orthodox]{nag}
\RequirePackage{snapshot}

\documentclass[9pt,onecolumn]{extarticle}

\makeatletter
\if@twocolumn
  \usepackage[dvips,letterpaper,top=0.5in, bottom=0.5in, left=0.75in, right=0.5in,includefoot,heightrounded]{geometry}
\else
  \usepackage[dvips,letterpaper,margin=1in,includefoot,heightrounded]{geometry}
\fi

\usepackage{srcltx}

\usepackage[russian,portuges,english]{babel}

\iflanguage{portuges}
    {\newcommand{\keywordname}{Palavras-chaves}}
    {\newcommand{\keywordname}{Keywords}}

\usepackage{amsmath}
\usepackage{amssymb,amsfonts}

\usepackage{abstract}

\usepackage{graphicx}
\usepackage[usenames,dvipsnames,svgnames,x11names]{xcolor}
\usepackage{color, colortbl}
\definecolor{LBlue}{rgb}{0.63, 0.79, 0.95}
\usepackage{subfigure}

\usepackage{booktabs}

\usepackage{setspace}
\usepackage{flushend}

\usepackage{dsfont}

\usepackage{cite}

\usepackage{hyperref}
\usepackage[normalem]{ulem}
\usepackage{bm}

\usepackage{enumerate}

\usepackage{multirow}
\usepackage[noend]{algpseudocode}

\usepackage{listings}

\usepackage[short,12hr]{datetime}
       \usepackage{fouriernc}

\newcommand{\barma}{${\beta}$ARMA}

\newcommand{\tablesize}{\fontsize{9}{11}\selectfont}

\makeatletter

\newcommand{\printtitle}{%
\makeatletter
\if@twocolumn

\twocolumn[%
  \maketitle
  \begin{onecolabstract}
    \myabstract
  \end{onecolabstract}
  \begin{center}
    \small
    \textbf{\keywordname}
    \\\medskip
    \mykeywords
  \end{center}
  \bigskip
]
\saythanks
\else
  \maketitle
  \begin{onecolabstract}
    \myabstract
  \end{onecolabstract}
  \begin{center}
    \small
    \textbf{\keywordname}
    \\\medskip
    \mykeywords
  \end{center}
  \bigskip
  \onehalfspacing
\fi
\makeatother
}

\author{%
B. G. Palm%
\thanks{Departamento de Engenharia Eletr\^onica e Computa\c{c}\~ao,
	Instituto Tecnol\'ogico de Aeron\'autica, Brazil
and
Signal Processing Group, Departamento de Estat\'istica, Universidade Federal Pernambuco, Brazil. Email: \url{brunagpalm@gmail.com}
}
\and
F. M. Bayer%
\thanks{Departamento de Estat\'istica and LACESM, Universidade Federal de Santa Maria, Brazil. Email: \url{bayer@ufsm.br}}
\and
R.~J.~Cintra%
\thanks{%
Signal Processing Group,
Universidade Federal Pernambuco, Brazil.
Email: \url{rjdsc@de.ufpe.br}}
}

\title{%
Signal
Detection
and Inference
Based on
the
Beta Binomial
Autoregressive Moving Average
Model}

\newcommand{\myabstract}{%
This paper
proposes
the beta binomial
autoregressive moving average model~(BBARMA)
for
modeling
quantized amplitude data
and
bounded count data.
The BBARMA model estimates the conditional mean
of a beta binomial distributed variable observed over the time
by a dynamic structure
including:
(i)~autoregressive and moving average terms;
(ii)~a set of regressors;
and
(iii)~a link function.
Besides introducing the new model,
we develop parameter estimation,
detection tools,
an
out-of-signal
forecasting scheme,
and
diagnostic measures.
In particular,
we provide closed-form expressions
for the
conditional score vector and
the conditional information matrix.
The proposed model
was submitted
to extensive Monte Carlo simulations
in order to
evaluate the
performance of the
conditional maximum likelihood
estimators
and of the proposed detector.
The derived detector outperforms
the usual
ARMA- and Gaussian-based detectors
for sinusoidal signal detection.
We also presented
an experiment for modeling and forecasting
the
monthly number of rainy days
in Recife, Brazil.
}

\newcommand{\mykeywords}{%
ARMA filter,
beta binomial distribution,
detection,
digital signal,
time series.
}

\date{}

\begin{document}

\printtitle

\section{Introduction}

Signal detection is a
fundamental task in the field
of signal processing,
being pivotal for decision
making
and
information extraction~\cite{Kay1998-2}.
Over the years,
several
detection methods
have been developed
assuming additive Gaussian noise~\cite{Kay1998-2,wang2019,wang2019b},
with
decision criteria
based on
continuous-time waveforms~\cite{Kay1998-2}.
In contrast,
real-world
data often
present
non-Gaussian
signals~\cite{al2002,liu2018}
and
the
Gaussianity assumption
may not be
enough
to
model several practical contexts,
as illustrated, for example, in~\cite{wang2019b,
wang2019,
jacques2010}.
Furthermore,
signal processing
systems
operate
under
quantized
discrete-time sampled data.
Quantized discrete-time
signals---here referred to as digital signals---constitute
a clear example
of non-Gaussian data
and
their
estimation and detection have been
attracting attention over the past years~\cite{
wang2019b,
wang2018}.
Quantized signals
represent  data
according to
a finite
number of discrete values
observed over time~(e.g.,
256 amplitude levels in
8-bit quantization)~\cite{
rabiner2007,oppenheim2010}.
Thus, the application of
Gaussian-based detectors
and hypothesis tests
to non-Gaussian processes may
lead to suboptimal
or
erroneous
detectors~\cite{schwartz1989}.
Although unsatisfactory~\cite{Cribari2010,Pinheiro2011},
a better approach is
to consider a data transformation that
maps the measured signal
from its
original distribution into
the Gaussian distribution~\cite{Cribari2010,Pinheiro2011}.
This method
might
offer
limited results
because the transformed data
should
be interpreted
in terms
of the
transformed
signal mean,
not
in terms of the
measured data mean~\cite{Cribari2010}.
In this context,
the beta binomial distribution
offers another
way
to model
quantized discrete-time
signals.
The beta binomial distribution
has been used
over the years
to model
bounded discrete
values~\cite{Lee1987,werner2015},
arising
as a natural candidate
for digital signals
modeling.
Nevertheless,
to the best of our
knowledge,
a time series model
based on
the beta binomial distribution
capable of
addressing the detection problem
for digital signal
is absent in the literature.

Besides quantized signal processing applications,
the derived model can be used
to fit any
bounded count data observed over the time,
such as
the number of rainy days per time interval~\cite{gouveia2018},
number of
defective products in one lot~\cite{njike2012},
or
number of hospital admissions~\cite{albarracin2018}.
In particular,
the monitoring of the number of rainy days per time interval,
either monthly or weekly,
is
gaining importance in discrete-time signal modeling
since it
influences
on human health,
climate changes,
constructed environments,
agriculture,
and
economy~\cite{gouveia2018}.
For example,
in~\cite{tian2016},
the number
of rainy days is
employed to
improve urban water demand forecasting;
in~\cite{lacombe2014},
to accurately characterize
future changes in
water resource availability in India;
and
in~\cite{ehelepola2015},
to analyze Dengue fever
incidence~\cite{ehelepola2015}.

Our goal is twofold.
First,
we introduce
a time series model
for
quantized amplitude data
and
bounded count data observed over the time,
which
estimates the mean
of  beta binomial
distributed
signals.
The sought model
consists of
autoregressive and moving average terms,
a set of regressors,
and a link function.
For the
proposed
beta binomial autoregressive
moving average~(BBARMA)
model,
we
introduce
parameter
estimation,
the conditional
observed information matrix,
an
out-of-signal
forecasting tool,
and
diagnostic measures.
Second,
we
present
a signal detector
based on
the asymptotic properties
of the
sought model
parameter estimators.
The proposed detector
is suitable
for identifying
the presence of
particular signals
from
beta binomial distributed
measured data.

The paper is organized as follows.
In Section~\ref{s:model}, we
provide the mathematical
formalism of the derived model
and discuss the conditional maximum likelihood estimation.
Section~\ref{s:large} shows the
proposed
detection theory,
presenting
the
conditional
observed information
matrix,
a hypothesis test,
and the implied
signal detector.
In Section~\ref{s:forecasting},
diagnostic
measures
and
a
forecasting tool
are discussed.
Section~\ref{s:vali}
presents
Monte Carlo simulations
for
evaluating
the
derived
conditional maximum likelihood
and
assessing
the
performance of
the proposed model
in digital signal detection.
Also,
an
out-of-signal forecasting investigation
of
bounded count data observed over the time
is discussed.
Finally,
Section~\ref{s:conclusion}
concludes
the work.

\section{The Proposed Model}
\label{s:model}

The beta binomial model
was proposed in~\cite{Skellam1948},
but the idea of this distribution
goes back to E.~Pearson~\cite{Pearson1925}.
The beta binomial distribution is a
composition of the beta and the binomial distribution,
where the variable of interest,~$Y$,
is a
random variable
with binomial distribution,
where the probability of success
follows a beta distribution~\cite{Forcina1988}.
The beta binomial probability
function~(pf)
is given by~\cite{Bibby2011}
\begin{align*}
p_Y(y; K, a, b) &=
\frac{\Gamma(a+b)
\Gamma(K+1)}{\Gamma(a)
\Gamma(b)\Gamma(y+1)\Gamma(K-y+1)}
\frac{\Gamma(a+y)\Gamma(K-y+b)}{\Gamma(K+a+b)},
\end{align*}
where $a,b$ are strictly positive numbers,
$\Gamma(z)
=
\int_0^\infty
t^{z-1}
e^{-t}
\operatorname{ d }  t$,
for~$z > 0$,
is
the gamma
function~\cite{abramowitz1972},
and~$K$ is a positive integer.
The quantity~$y = 0, 1, 2, \ldots, K$
can be interpreted as
an observed signal value
and~$K$, as its maximum value.
The mean and the variance of~$Y$ are, respectively,
\begin{align*}
\operatorname{E}(Y) &=  \frac{a}{a+b} K
,
\\
\operatorname{Var}(Y) &=
\frac{ab}{(a+b)^2(a+b+1)}
\left( K^2 + (a+b)K \right)
,
\end{align*}
where~$\mu = a/(a+b) $
which
can be understood
as the mean
of~$Y/K$.
In the next section,
we introduce
a dynamic model
to fit the mean parameter.

\subsection{Time Series Model}

In order to
define the BBARMA model,
we consider a new
parameterization in the
beta binomial distribution
based on
(i)~a precision parameter,~$\varphi$,
and
(ii)~the quantity $\mu$.
The proposed parameters
satisfy the following relations:
$a = \mu \varphi$
and~$b= (1- \mu ) \varphi$.

Let~$\{Y[n]\}_{n\in \mathds{Z}}$
be a stochastic process,
where each~$Y[n]$
assumes values~$y[n]$ between
zero and~$K$.
Let~$\mathcal{F} [n]$
be the sigma-field generated by
past observations
$\lbrace \ldots ,
y[n-2], y[n-1], y[n]
\rbrace$.
Assume that,
conditionally to the previous information
set~$\mathcal{F}[n-1]$,
each~$Y[n]$ is distributed according to
the
beta binomial distribution
with parameters~$\mu[n]$ and~$\varphi$,
where~$\mu[n]$ is the conditional mean
of~$Y [n] / K$.
The conditional probability function of~$Y[n]$
given~$\mathcal{F}[n-1]$
is defined as
\begin{align*}
f_Y(y[n]\mid\mathcal{F}[n-1])=
\frac{\Gamma(K+1)}{\Gamma(y[n]+1)\Gamma(K-y[n]+1)}
\frac{\Gamma(\varphi)\Gamma(y[n]+\mu[n] \varphi)
\Gamma(K-y[n]+(1-\mu[n])\varphi)}{\Gamma(K+\varphi)
\Gamma(\mu[n] \varphi) \Gamma((1-\mu[n])\varphi)}
.
\end{align*}
The conditional mean and conditional
variance of~$Y[n]$,
given~$\mathcal{F}[n-1]$,
are
respectively given by
\begin{align*}
\operatorname{E}(Y[n]\mid \mathcal{F}[n-1])
&= \mu[n] K,
\\
\operatorname{Var}(Y[n] \mid \mathcal{F}[n-1])
&=  (\mu[n]-\mu[n]^2) K
\frac{K + \varphi}{1+\varphi}
.
\end{align*}

\begin{figure}
\centering
\subfigure[$\;\varphi=4$]
{\label{f:densitys_10}
	\includegraphics[width=0.48\textwidth]{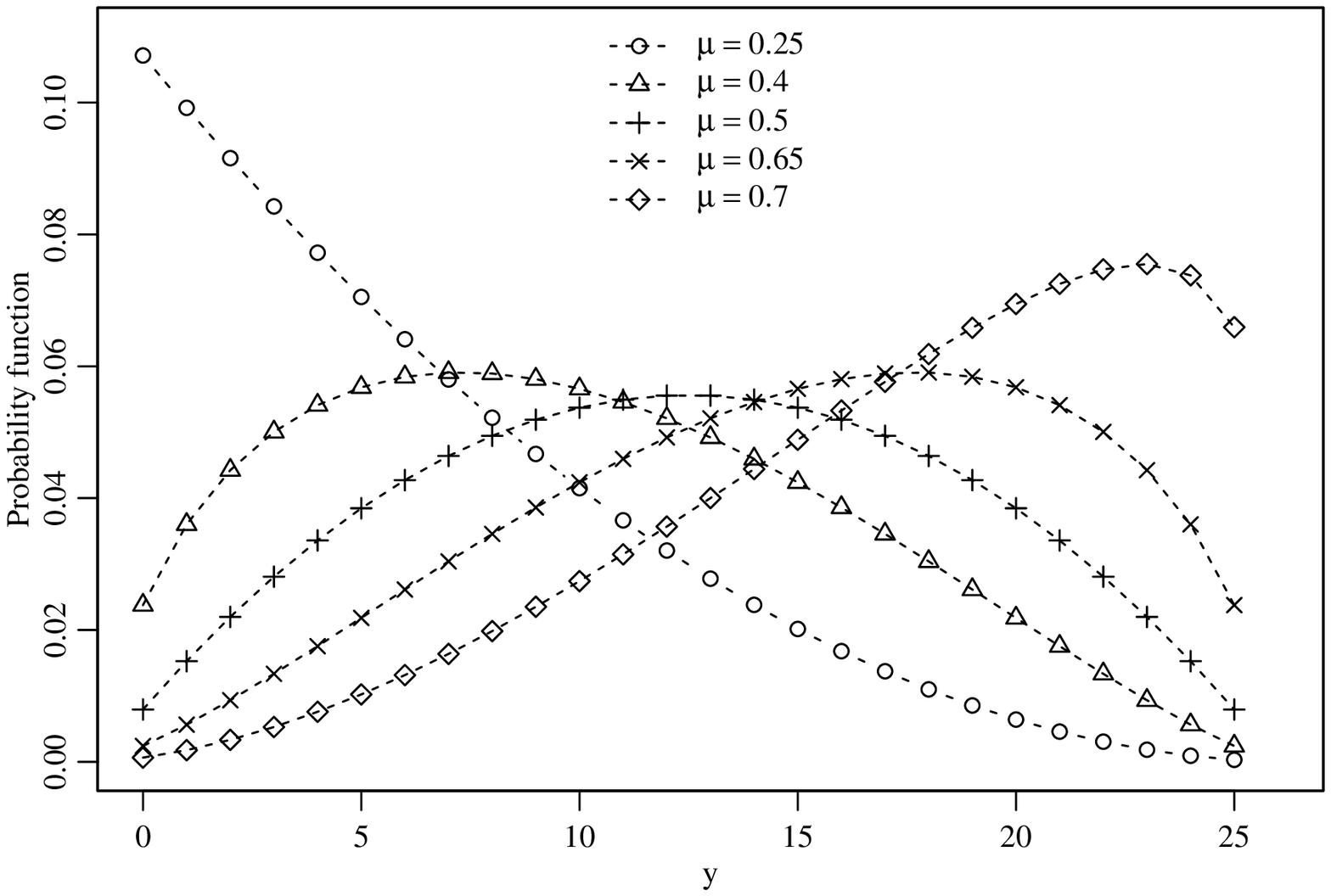}}
\subfigure[$\;\varphi=100$]
{\label{f:densitys_90}
	\includegraphics[width=0.48\textwidth] {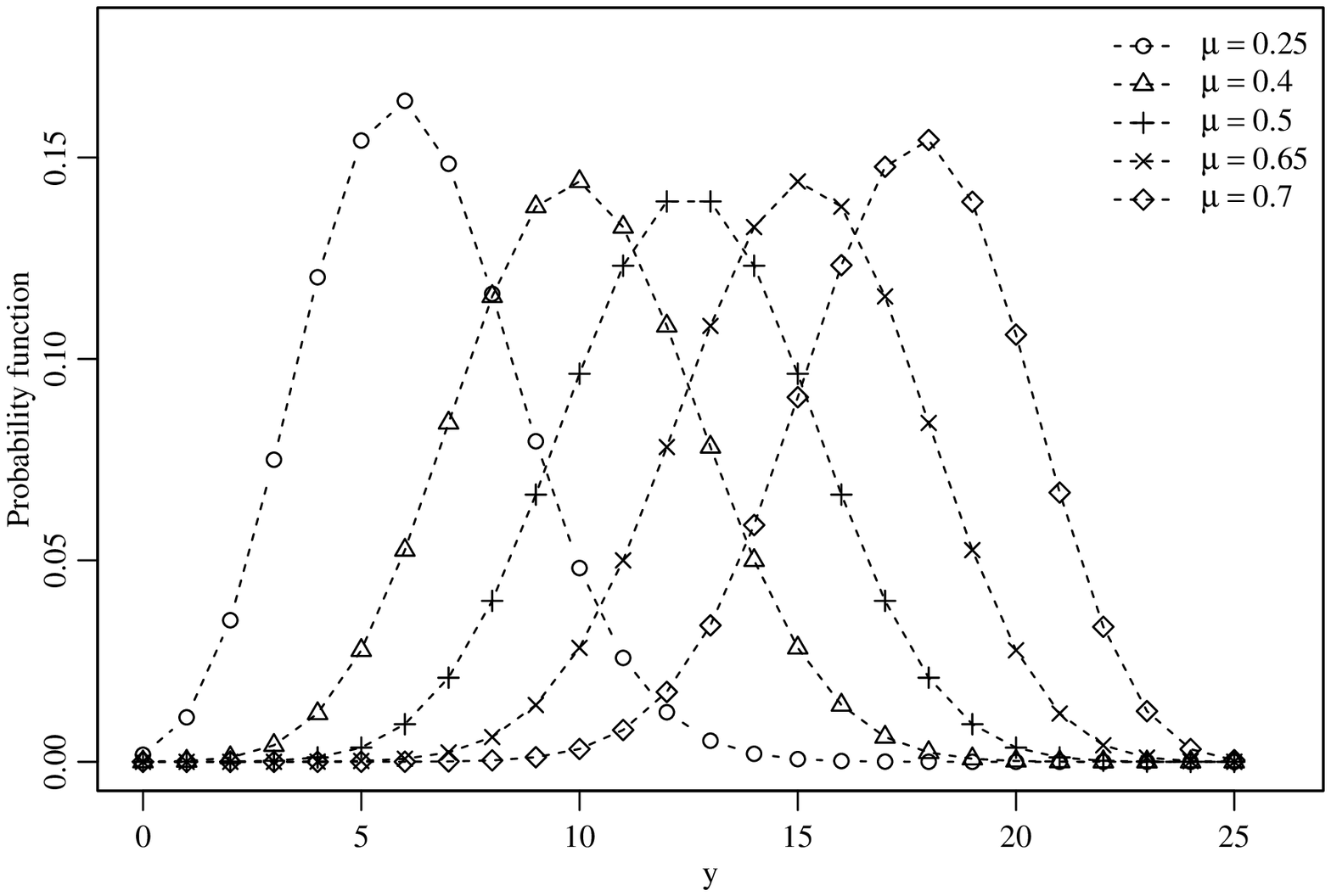}}
\caption{
Beta binomial probability functions for different values
of~$\mu$,
$K=25$,
and (a)~$\varphi = 4$, (b)~$\varphi = 100$.
}\label{f:fig}
\end{figure}

The beta binomial probability
function is very flexible,
as shown in Figure~\ref{f:fig}.
For
small values of~$\varphi$,
the beta binomial
distribution
can present
decreasing,
increasing-decreasing,
and
upside-down bathtub
shapes.
On the other hand,
for
large values of $\varphi$,
the beta binomial
distribution
accommodates more symmetric distributions,
i.e.,
data around the mean.

The
BBARMA model
relates the mean~$\mu[n]$ to one
linear predictor,~$\eta[n]$~\cite{McCullagh1989}.
This relation is based on a
strictly monotonic
and
twice differentiable
function,~$g(\cdot)$, called
link function~\cite{McCullagh1989}.
Popular
choices
for the
link functions
are
the logit~\cite{linkfunction},
the probit~\cite{McCullagh1989},
and
the complementary log-log~\cite{linkfunction, Benjamin2003, Rocha2009}.
Following the generalized linear models~(GLM)
mathematical formalism~\cite{McCullagh1989},
we have that
\begin{align*}
g(\mu [n] ) =
\eta[n]
=\mathbf{x}^\top[n] \bm{\beta},
\end{align*}
where~$\bm{\beta} =
(\beta_1, \beta_2, \ldots, \beta_l)^\top$
is the set of unknown
parameters
and
$\mathbf{x}[n]= (x_1[n], x_2[n], \ldots, x_l[n])^\top$,
$n=1,2,\ldots, N$,
is the vector of the covariates with~$l <N$.

Therefore,
similar to
the generalized autoregressive moving average~(GARMA)
model~\cite{Benjamin2003}
and
the beta autoregressive moving average~(\barma)
model~\cite{Rocha2009,rocha2017},
a
dynamical
general model for~$\mu[n]$
is given by~\cite{Benjamin2003}
\begin{align*}
g(\mu[n]) = \eta[n] =\mathbf{x}^\top [n]
\bm{\beta}  + \bm{\tau}[n],
\end{align*}
where
\begin{align}
\begin{split}
\label{e:modelogeral}
\bm{\tau}[n] =&
\sum \limits_{i=1}^p
\phi_i \mathbf{\mathcal{A}}
(y[n-i])
+ \sum \limits _{j=1}^q \theta_j
\mathbf{\mathcal{M}} (y[n-j],\mu[n-j]),
\end{split}
\end{align}
where
$p$
and~$q$ are the orders of the model;
and the
autoregressive and moving average terms
are
represented by the functions~$\mathbf{\mathcal{A} (\cdot)}$
and~$\mathbf{\mathcal{M} (\cdot)}$,
respectively.
The quantities~$\phi_i$,~$i=1,2,\ldots,p$,
and~$\theta_j$,~$j=1,2,\ldots,q$,
are
the autoregressive and moving average parameters,
respectively.
By
including an
intercept~$ \zeta \in \mathbb{R}$,
we extend the model
described
in~\eqref{e:modelogeral},
yielding:
\begin{align}
\label{model}
\begin{split}
g(\mu[n]) &= \eta[n] =
\zeta +
\mathbf{x} ^\top [n] \bm{\beta}
+ \sum \limits _{i=1} ^p \phi_i
 y^\star [n-i]
+ \sum \limits _{j=1} ^q \theta_j \{ y^\star [n-j] - \mu[n-j]\},
\end{split}
\end{align}
where~$y^\star [n] = y [n] / K$
is the scaled observed signal
to ensure its
value
at the same range
of~$\mu [n]$.

\subsection{Conditional Maximum Likelihood Estimation}
\label{s:estimation}

The estimation of BBARMA$(p,q)$ model parameters
can be
realized by maximizing the
logarithm of the conditional
likelihood function~\cite{brockwell2016,Benjamin2003}.
Let the vector parameters be
$\bm{\gamma}=(\zeta,\bm{\beta}^\top,
\bm{\phi}^\top,\bm{\theta}^\top,\varphi)^\top$,
with
$\bm{\beta} = (\beta_1, \beta_2,\ldots,
\beta_l)^\top$,
$\bm{\phi} = (\phi_1, \phi_2,
\ldots, \phi_p)^\top$,
and
$\bm{\theta} = (\theta_1, \theta_2,
\ldots, \theta_q)^\top$.
Based on an observed signal of length $N$,~$(y[1], y[2], \ldots , y[N])$,
the
log-likelihood function for the
vector
parameter~$\bm{\gamma}$ conditional to
the~$m=\max(p,q)$ preliminary observations is given by
\begin{align}
\label{log}
\ell=
\sum_{n=m+1}^{N} \log f(y[n]\mid\mathcal{F}[n-1])
= \sum_{n=m+1}^{N} \ell[n] (\mu[n],\varphi),
\end{align}
where
\begin{align*}
\ell[n](\mu[n], \varphi) =&
\log \Gamma(K+1) -
\log \Gamma (y[n]+1)
- \log \Gamma(K-y[n]+1)
- \log \Gamma ({\mu[n]}{\varphi})
\\&
+ \log \Gamma (y[n]+{\mu[n] }{\varphi})
-\log \Gamma (K+\varphi )
+ \log \Gamma (K-y[n]+
({1-\mu[n]}){\varphi})
\\ &
 -
 \log \Gamma (({1-\mu[n]}){\varphi})
 +
 \log \Gamma (\varphi)  .
\end{align*}
The
conditional maximum likelihood estimators~(CMLE),
$\widehat{\bm{\gamma}}$,
can be obtained
from the score vector~$\mathbf{U}(\bm{\gamma})$
by solving
\begin{align}
\label{e:escore}
\mathbf{U}(\bm{\gamma}) =
\frac{\partial \ell}{\partial \bm{\gamma}^\top} =
\left( \frac{\partial \ell }{\partial \zeta},
\frac{\partial \ell } {\partial \bm{\beta}^\top},
\frac{\partial \ell }{\partial \bm{ \phi}^\top},
\frac{\partial \ell }{\partial \bm{\theta}^\top},
\frac{\partial \ell }{\partial \varphi} \right)^\top
=
\bm{0}
,
\end{align}
where~$\bm{0}$
is a vector of zeros
with dimension~$p+q+l+2$.
By the chain rule,
for $\gamma_j \neq \varphi$,
$j=1, 2 , \ldots , p+q+l+1$,
we have
\begin{align*}
\frac{\partial \ell}{\partial
\gamma_j} =
\sum \limits _{n=m+1} ^{N}
\frac{\partial \ell[n](\mu[n], \varphi) }{\partial \mu[n]}
\frac{\operatorname{ d }  \mu[n]}{\operatorname{ d } \eta[n]}
\frac{\partial \eta[n]}{\partial
\gamma_j}
.
\end{align*}
Note that
\begin{align*}
\frac{\partial \ell[n] (\mu[n],\varphi)}
{\partial \mu[n]}
=&
\varphi \Upsilon[n]
,
\end{align*}
where
\begin{align*}
\Upsilon[n]
=&
\varphi \left[ \psi \left(y[n] + {\mu[n]}
{\varphi}\right)
- \psi \left(K-y[n]+({1-\mu[n]})
{\varphi}\right)
- \psi \left({\mu[n]}
{ \varphi}\right)
+ \psi \left(({1-\mu[n]})
{\varphi}\right) \right]
,
\end{align*}
and
\begin{align*}
\frac{\operatorname{ d } \mu[n]}
{ \operatorname{ d } \eta[n]} =& \frac{1} {g^\prime (\mu[n])}
,
\end{align*}
where~$\psi (\cdot)$ is the digamma function, i.e.,
$\psi (z)  =
\frac{\operatorname{ d } \log \Gamma (z)}{\operatorname{ d }z }
$,
for~$z >0$,
and~$g^\prime(\cdot)$
is the first derivative of the adopted link function~$g(\cdot)$.
In particular, for the logit link function,
$g(\mu[n]) = \log \left( \frac{\mu[n]}{1-\mu[n]} \right)$,
we have
$g^\prime(\mu[n]) =
\left( \mu[n] (1-\mu[n]) \right) ^{-1}
$.

Additionally,
similar to the \barma~model~\cite{rocha2017},
we have
\begin{align*}
\frac{\partial \eta [n]}{\partial \zeta}
&=
1
-
\sum \limits _{s=1}^q \theta_s
\frac{1}{g^\prime ( \mu [n-s])}
\frac{\partial \eta[n-s]}{\partial \zeta}
,
\\
\frac{\partial \eta [n]}{\partial  \beta_k}
&=
\mathbf{x} ^\top [n]
-
\sum \limits _{s=1}^q \theta_s
\frac{1}{g^\prime ( \mu [n-s])}
\frac{\partial \eta[n-s]}{\partial \beta_k}
,
\\
\frac{\partial \eta [n]}{\partial  \phi_i}
&=
y^\star[n-i]
- \sum \limits _{s=1}^q \theta_s
\frac{1}{g^\prime ( \mu [n-s])}
\frac{\partial \eta [n-s]}
{\partial \phi_i}
,
\\
\frac{\partial \eta [n]}{\partial  \theta_j}
&=
y^\star [n-j] - \mu[n-j]
- \sum \limits _{s=1}^q \theta_s
\frac{1}{g^\prime ( \mu [n-s])}
\frac{\partial \eta[n-s]}
{\partial \theta_j}
,
\end{align*}
for~$k=1,2, \ldots , l$,~$i=1,2,\ldots,p$,
and~$j=1,2,\ldots,q$.

The score function with respect to parameter~$\varphi$
is given by
\begin{align*}
\frac{\partial \ell}{\partial \varphi}
= &
\sum \limits _{n=m+1} ^{N}
\mu[n] \left[ \psi\left(y[n] + {\mu[n]}{\varphi}\right)
- \psi\left({\mu[n]}{\varphi}\right) \right]
+ (1-\mu[n]) \left[ \psi\left(K-y[n] +
({1-\mu[n]}){\varphi}\right)
\right. \\& \left.
 -\psi \left(({1-\mu[n]}){\varphi}\right) \right]
 +
 \psi \left(\varphi \right)
-\psi\left(K+\varphi\right)
 .
\end{align*}

The solution of~\eqref{e:escore}
has no
closed-form, thus, nonlinear
optimization algorithms,
such as Newton or
quasi-Newton~\cite{nocedal1999},
are necessary.
We selected the
Broyden-Fletcher-Goldfarb-Shanno~(BFGS)
method~\cite{press}
with analytic first derivatives,
due to its
superior performance for
non-linear
optimization method
requiring
only the
first derivatives~\cite{press}.
The
initial values for
the constant~($\zeta$),
the autoregressive~($\bm{\phi}$) parameters,
and the regressors~($\bm{\beta}$)
were
derived
from the ordinary least squares estimate
associated to the linear regression.
The response vector
is
$(y^\star[m+1], y^\star[m+2],
\ldots, y^\star[N])^\top$
and
the covariate matrix is given by
\begin{align*}
\begin{bmatrix}
1 & x_1[m] &
x_2[m]
& \ldots & x_l[m]
& y^\ast[m] &
y^\ast[m-1]
& \cdots & y^\ast[m-p+1] \\
1 & x_1[m+1] & x_2[m+1] &
\ldots
& x_l[m+1]
& y^\ast[m+1]
& y^\ast[m]
& \cdots & y^\ast[m-p+2] \\
\vdots %
& \vdots
& \vdots
& \ddots & \vdots & \vdots & \vdots & \ddots & \vdots \\
1 & x_1[N]
& x_2[N]
& \ldots
& x_l[N]
& y^\ast[N-1]
& y^\ast[N-2]
& \cdots & y^\ast[N-p] \\
\end{bmatrix}.
\end{align*}
For
the initial values,
we adopted~$\bm{\theta} = \bm{0}$,
as
suggested
in~\cite{Palm2017,bayer2017}
and~$\varphi=1$,
following~\cite{gamlss.dist}.

\section{Signal Detection Theory}
\label{s:large}

The goal of this section
is to introduce
a detector
based on
a binary hypothesis test
tailored for the BBARMA model.
For such,
we derive
the
conditional observed
information matrix
and
the
asymptotic properties
of the
conditional
parameter estimators
of the BBARMA model.

\subsection{Conditional Observed Information Matrix}
\label{s:fisher}

The expected and observed information
matrices are not identical
but they are asymptotically equivalent~\cite{Pawitan2001},
being the latter preferable for
hypothesis
testing when
the
distribution is not
in the exponential family~\cite{Efron1978,Pawitan2001}.
The
conditional
observed
information
matrix
is given by
the negative of the second order derivatives
of the conditional log-likelihood,
which
is
given by
\begin{align}
\label{e:fisher}
\begin{split}
\frac{\partial^2 \ell}{\partial \bm{\lambda} \partial \bm{\delta}}
=&
\sum \limits_{n=m+1}^N
\frac{\partial}{\partial\bm{\lambda}}
\left(
\frac{\partial \ell[n] (\mu[n],\varphi)}
{\partial \mu[n]}
\frac{\operatorname{ d } \mu[n]}{\operatorname{ d } \eta[n]}
\frac{\partial \eta[n]}{\partial \bm{\delta}}
\right)
\\
=&
\sum \limits_{n=m+1}^N
\left[
\frac{\partial^2 \ell[n](\mu[n], \varphi)}
{\partial \mu[n]^2}
\left(
\frac{\operatorname{ d } \mu[n]}{\operatorname{ d } \eta[n]}
\right)^2
\frac{\partial \eta[n]}{\partial \bm{\delta}}
\frac{\partial \eta[n]}{\partial \bm{\lambda}}
+
\frac{\partial \ell[n](\mu[n], \varphi)}{\partial \mu[n]}
\frac{\operatorname{ d }^2 \mu[n]}{\operatorname{ d } \eta[n]^2}
\frac{\partial \eta[n]}{\partial \bm{\delta}}
\frac{\partial \eta[n]}{\partial \bm{\lambda}}
\right.
\\&
\left.
+
\frac{\partial \ell[n](\mu[n], \varphi)}{\partial \mu[n]}
\frac{\operatorname{ d } \mu[n]}{\operatorname{ d } \eta[n]}
\frac{\partial^2 \eta[n]}{\partial \bm{\delta} \partial \bm{\lambda}}
\right]
,
\end{split}
\end{align}
where
$\bm{\lambda} = (\zeta,\bm{\beta}^\top,\bm{\phi}^\top,
\bm{\theta}^\top)^\top$
and
$\bm{\delta} = (\zeta,\bm{\beta}^\top,\bm{\phi}^\top,
\bm{\theta}^\top)^\top$.
The derivatives
$\frac{\partial \ell[n](\mu[n], \varphi)}
{\partial \mu[n]}$,
$\frac{\operatorname{ d } \mu[n]}
{\operatorname{ d } \eta[n]}$,
$\frac{\partial \eta[n]}
{\partial \bm{\delta}}$,
and
$\frac{\partial \eta[n]}
{\partial \bm{\lambda}}$
are given in Section~\ref{s:estimation}.
The
second order derivative
of $\ell[n](\mu[n],\varphi)$ with respect to $\mu[n]$
is given by
\begin{align*}
\frac{\partial^2 \ell[n] (\mu[n],\varphi)}{\partial \mu[n]^2} =&
\varphi^2
\big\lbrace
\psi^\prime \left( y[n] + {\mu[n]}{\varphi}\right)
+  \psi^\prime \left( K - y[n]
 + ({1 - \mu[n]}){\varphi}\right)
 \\&
- \psi^\prime \left( ({1 -\mu[n]}){\varphi}\right)
-\psi^\prime \left({\mu[n]}{\varphi}\right)
\big\rbrace,
\end{align*}
where~$\psi^\prime (z)  =
\frac{\operatorname{ d } \psi (z)}{\operatorname{ d }z }
$,
for~$z >0$,
is the first derivative
of the digamma function,
i.e.,
the trigamma function~\cite{abramowitz1972}.
Note that
\begin{align*}
\frac{\operatorname{ d }^2 \mu[n]}{\operatorname{ d } \eta[n]^2}
=
-\frac{g^{\prime \prime} (\mu[n])}{\left(g^\prime(\mu[n])\right)^2}
.
\end{align*}
Additionally,
the second order
derivative of~$\frac{\partial \eta[n]}
{\partial \bm{\delta}}$
with respect
to~$\zeta$,~$\phi$,
and~$\beta$,
for~$\delta_i \neq \theta_j$,
where~$i = 1, 2, \ldots ,(p+l+1)$
and~$j=1,2,\ldots,q$,
is equal to zero.
On the other hand,
\begin{align*}
\frac{\partial^2 \eta[n]}
{\partial \bm{\delta} \partial \theta_j}
=&
-
\frac{d \mu [n-j]}{d \eta [n-j]}
\frac{\partial \eta [n-j]}{\partial \bm{\delta}}
-
\sum \limits_{s=1}^q
\theta_s
\frac{\partial \eta [n-s]}{\partial \bm{\delta}}
\frac{d^2 \mu [n-s]}{d \eta [n-s]^2}
\frac{\partial \eta [n-s]}{\partial \theta_j}
-
\sum \limits_{s=1}^q
\theta_s
\frac{d \mu [n-s]}{d \eta [n-s]}
\\&
\cdot
\frac{\partial^2 \eta [n-s]}
{\partial \bm{\delta} \partial \theta_j}
,
\\
\frac{\partial^2 \eta[n]}
{\partial \theta_k \partial \theta _ j}
=&
-
\frac{d \mu [n-j]}{d \eta [n-j]}
\frac{\partial \eta [n-j]}{\partial \theta_k}
-
\frac{d \mu [n-k]}{d \eta [n-k]}
\frac{\partial \eta [n-k]}{\partial \theta_j}
-
\sum \limits_{s=1}^q
\theta_s
\frac{\partial \eta [n-s]}{\partial \theta_j}
\frac{d^2 \mu [n-s]}{d \eta [n-s]^2}
\frac{\partial \eta [n-s]}{\partial \theta_k}
\\&
-
\sum \limits_{s=1}^q
\theta_s
\frac{d \mu [n-s]}{d \eta [n-s]}
\frac{\partial^2 \eta [n-s]}
{\partial \theta_k \partial \theta_j}
,
\end{align*}
for~$j,k=1,2,\ldots,q$.

The derivative
of
$\frac{\partial \ell}{\partial  \bm{\delta}}$
with respect to~$\varphi$
is given by
\begin{align*}
\frac{\partial^2 \ell}{\partial \bm{\delta} \partial \varphi}
=
\sum \limits_{n=m+1}^N
\frac{\partial \eta [n]}{\partial \bm{\delta} }
\frac{1}{g^\prime (\mu[n]) }
\frac{\partial \ell[n] (\mu[n],\varphi) }{\partial \mu[n] \partial \varphi}
,
\end{align*}
where
\begin{align*}
\frac{\partial \ell[n] (\mu[n],\varphi) }{\partial \mu[n] \partial \varphi}
=&
\varphi
\left\lbrace
\left(1-\mu[n]\right)
\left[
{\psi^\prime \left(({1-\mu[n]}){\varphi}\right)}
-\psi^\prime \left(K-y[n]
+({1-\mu[n]}){\varphi}\right)
\right]
\right.  \\& \left.
+\mu[n]
\left[
\psi^\prime \left(y[n]+{\mu[n]}{\varphi}\right)
-\psi^\prime \left( {\mu[n]}{\varphi}\right)
\right]
\right \rbrace
+ \Upsilon[n]
.
\end{align*}
The
second order
derivative
of~$\ell[n](\mu[n],\varphi)$ with respect to~$\varphi$
is given by
\begin{align*}
\frac{\partial^2 \ell[n] (\mu[n],\varphi)}{\partial \varphi^2} =&
\psi^\prime (\varphi) + \mu[n]^2
\left[
\psi^\prime
(y[n] +\mu[n] \varphi)
- \psi^\prime (\mu[n] \varphi) \right]
+ (1-\mu[n])^2
\left[
\psi^\prime (K-y[n]
\right. \\& \left.
+
(1-\mu[n])\varphi)
- \psi^\prime ((1-\mu[n])\varphi)
\right]
.
\end{align*}
To facilitate the
presentation
of the
conditional
observed
information matrix,
we introduce
the following
auxiliary
vectors and
matrices.
Let~$\mathbf{1}$ be the~$(N-m) \times 1 $ vector of ones,
$\mathbf{T} =
\operatorname{diag} \lbrace 1/g^\prime (\mu[m+1]),
1/g^\prime (\mu[m+2]), \ldots, 1/g^\prime (\mu[N])\rbrace$,
$\mathbf{a} = \left( \frac{\partial \eta[m+1]}{\partial \zeta},
\frac{\partial \eta[m+2]}{\partial \zeta},
\ldots
,
\frac{\partial \eta[N]}{\partial \zeta}
\right)^\top$,
$
\bm{\Upsilon} =
\left( \Upsilon[m+1],
\Upsilon[m+2], \ldots, \Upsilon[N] \right)^\top
$,
$
\Upsilon^\ast [n] =
\frac{\partial^2 \ell[n] (\mu[n],\varphi)}
{\partial \mu[n]^2}
$,
$
\Upsilon_\varphi [n] =
\frac{\partial^2 \ell[n] (\mu[n],\varphi)}
{\partial \mu[n] \partial \varphi}
$,
$
\Phi^\ast[n] =
\frac{\partial^2  \ell[n](\mu[n], \varphi)}
{ \partial \varphi^2}
$,
$\kappa[n] = \frac{ g^{\prime \prime}(\mu[n]) }
{ (g^\prime (\mu[n]))^2 }$,
$\xi[n] = \varphi
\kappa[n]
\Upsilon[n]
-\Upsilon^\ast[n]
\mathbf{T}^2
$,
$\mathbf{A} = \left( \frac{\partial^2 \eta[m+1]}
{\partial \zeta \partial \bm{\theta}},
\frac{\partial^2 \eta[m+2]}
{\partial \zeta \partial \bm{\theta}},
\ldots
,
\frac{\partial^2 \eta[N]}
{\partial \zeta \partial \bm{\theta}}
\right)^\top$,
$\mathcal{W}
=
\left(
\varphi
\Upsilon[m+1]
\mathbf{T}
,
\varphi
\Upsilon[m+2]
\mathbf{T}
,
\ldots
,
\varphi
\Upsilon[N]
\mathbf{T}
\right)
$,
$
\mathbf{L}
=
\operatorname{diag} \left \lbrace \Phi^\ast [m+1],
\Phi^\ast [m+2], \ldots , \Phi^\ast [N] \right \rbrace
$,
$
\mathbf{D}
=
\operatorname{diag} \left \lbrace \Upsilon_\varphi [m+1],
\Upsilon_\varphi [m+2], \ldots ,
\Upsilon_\varphi [N] \right \rbrace
$,
$
\mathbf{W}
=
\operatorname{diag} \left \lbrace
\xi[m+1],
\xi[m+2],
\ldots ,
\xi[N]
\right \rbrace
$,
$
\mathbf{M}[i,j] = \frac{\partial \eta [i+m]}{\partial \beta_j}
$,
$
\mathbf{P}[i,j] = \frac{\partial \eta [i+m]}{\partial \phi_j}
$,
$
\mathbf{R}[i,j] = \frac{\partial \eta [i+m]}{\partial \theta_j}
$,
$
\mathcal{M}[i,j] = \frac{\partial^2 \eta [i+m]}
{\partial \beta_i \partial \theta_j}
$,
$
\mathcal{P}[i,j] = \frac{\partial^2 \eta [i+m]}
{\partial \phi_i \partial \theta_j}
$,
and
$
\mathcal{R}[i,j] = \frac{\partial^2 \eta [i+m]}
{\partial \theta_i \partial \theta_j}
$.
The matrices~$\mathbf{M}[\cdot,\cdot]$,
$\mathbf{P}[\cdot,\cdot]$,
$\mathbf{R}[\cdot,\cdot]$,
$\mathcal{M}[\cdot,\cdot]$,
$\mathcal{P}[\cdot,\cdot]$,
and~$\mathcal{R}[\cdot,\cdot]$
are of dimensions~$(N-m) \times l$,
$(N - m) \times p$,
$(N - m) \times q$,
$(N-m) \times l$,
$(N - m) \times p$,
and~$(N - m) \times q$,
respectively.
Based on~\eqref{e:fisher}
and on the above expressions,
the
conditional
observed information matrix is
given by
\[
\mathbf{I}(\bm{\gamma}) =
\setlength\arraycolsep{1pt}
\left[ \begin{array}{ccccc}
I_{(\zeta,\zeta)} &
\mathbf{I}_{(\zeta,\beta)} &
\mathbf{I}_{(\zeta,\phi)} &
\mathbf{I}_{(\zeta,\theta)} &
\mathbf{I}_{(\zeta,\varphi)}
\\
\mathbf{I}_{(\beta,\zeta)} &
\mathbf{I}_{(\beta,\beta)} &
\mathbf{I}_{(\beta,\phi)} &
\mathbf{I}_{(\beta,\theta)} &
\mathbf{I}_{(\beta,\varphi)}
\\
\mathbf{I}_{(\phi,\zeta)} &
\mathbf{I}_{(\phi,\beta)} &
\mathbf{I}_{(\phi,\phi)} &
\mathbf{I}_{(\phi,\theta)} &
\mathbf{I}_{(\phi,\varphi)}\\
\mathbf{I}_{(\theta,\zeta)} &
\mathbf{I}_{(\theta,\beta)} &
\mathbf{I}_{(\theta,\phi)} &
\mathbf{I}_{(\theta,\theta)} &
\mathbf{I}_{(\theta,\varphi)}
\\
\mathbf{I}_{(\varphi,\zeta)} &
\mathbf{I}_{(\varphi,\beta)} &
\mathbf{I}_{(\varphi,\phi)} &
\mathbf{I}_{(\varphi,\theta)} &
I_{(\varphi,\varphi)}
\\
\end{array} \right]
,
\]
where
$ I_{(\zeta,\zeta)} = \mathbf{a^\top W a}$,
$ \mathbf{I}_{(\zeta,\beta)} =
\mathbf{I}_{(\beta,\zeta)} ^\top =
\mathbf{M^\top W a}  $,
$\mathbf{I}_{(\zeta,\phi)}  =
\mathbf{I}_{(\phi,\zeta)} ^\top =
\mathbf{P^\top W a}$,
$\mathbf{I}_{(\zeta,\theta)}  =
\mathbf{I}_{(\theta,\zeta)}^\top =
\mathbf{R^\top W a}
-
\mathcal{W}
\mathbf{A}
$,
$\mathbf{I}_{(\zeta,\varphi)} =
\mathbf{I}_{(\varphi,\zeta)} ^\top =
\mathbf{-a^\top T D \mathbf{1}} $,
$\mathbf{I}_{(\beta,\beta)} =
\mathbf{M^\top W M}$,
$\mathbf{I}_{(\beta,\phi)} =
\mathbf{I}_{(\phi,\beta)} ^\top =
\mathbf{P^\top W M}$,
$\mathbf{I}_{(\beta,\theta)}  =
\mathbf{I}_{(\theta,\beta)}^\top =
\mathbf{R^\top W M}
-
\mathcal{W} \mathcal{M}$,
$\mathbf{I}_{(\beta,\varphi)} =
\mathbf{I}_{(\varphi,\beta)}^\top  =
\mathbf{-M^\top T D \mathbf{1}}$,
$\mathbf{I}_{(\phi,\phi)} =
\mathbf{P^\top W P} $,
$\mathbf{I}_{(\phi,\theta)} =
\mathbf{I}_{(\theta,\phi)}^\top =
\mathbf{R^\top W P}
-
\mathcal{W P}$,
$\mathbf{I}_{(\phi,\varphi)} =
\mathbf{I}_{(\varphi,\phi)}^\top =
\mathbf{-P^\top T D \mathbf{1}}$,
$\mathbf{I}_{(\theta,\theta)} =
\mathbf{R^\top W R}
-
\mathcal{W R}$,
$\mathbf{I}_{(\theta,\varphi)} =
\mathbf{I}_{(\varphi,\theta)}^\top =
\mathbf{ -R^\top T D \mathbf{1}} $,
and
$I_{(\varphi,\varphi)} = -\operatorname{tr}(\mathbf{L})$,
where~$\operatorname{tr}(\cdot)$ is the trace function.

Based on the consistency of the CMLE
and on
the asymptotic
distribution
of~$\widehat{\bm{\gamma}}$,
which is
given by~\cite{Kay1993,Pawitan2001},
we have that
\begin{align*}
\sqrt{N}
(\widehat{\bm{\gamma}}-\bm{\gamma})
\stackrel[{N \rightarrow \infty}]{d}{\longrightarrow}
\mathcal{\mathbf{N}} \left(\bm{0},
\mathbf{I}^{-1}(\bm{{\gamma}})\right)
,
\end{align*}
where~$\stackrel{d}{\longrightarrow} $
denotes convergence in distribution
and~$\mathcal{\mathbf{N}}$
is the multivariate
Gaussian distribution
with
null mean
and
covariance matrix~$\mathbf{I}^{-1}(\bm{{\gamma}})$.

\subsection{Hypothesis Test}

Let the
parameter vector~$\bm{\gamma}$
be
partitioned
in a parameter vector
of interest~$\bm{\gamma}_I$,
of dimension~$\nu$,
and
a vector of nuisance parameters,~$\bm{\gamma}_J$,
of dimension~$r-\nu$,
$r=1,2,\ldots, p+q+l+2$~\cite{Kay1998-2}.
In addition,~$\mathcal{H}_0:\bm{\gamma}_{I}=\bm{\gamma}_{I_0}$
is the hypothesis of interest
and~$\mathcal{H}_1:\bm{\gamma}_{I} \neq \bm{\gamma}_{I_0}$
the alternative hypothesis.
We selected
the
Wald statistic
as suggested in~\cite{Cribari2010},
which
is given by~\cite{Kay1998-2}
\begin{align*}
\\
T_W
&=
(\widehat{\bm{\gamma}}_{I_1}-\bm{\gamma}_{I_0})^\top
\left( \left[ \mathbf{I} ^ {-1}
(\widehat{\bm{\gamma}}_1) \right] _{\gamma_I \gamma _I} \right)^{-1}
(\widehat{\bm{\gamma}}_{I_1}-\bm{\gamma}_{I_0})
,
\end{align*}
where
$\widehat{\bm{\gamma}}_1
= (\widehat{\bm{\gamma}}_{I_1}^\top,
\widehat{\bm{\gamma}}_{J_1}^\top)^\top$
is the CMLE under~$\mathcal{H}_1$
and
$
\left[ \mathbf{I}^{-1} (\widehat{\bm{\gamma}})
\right]_{\gamma_I \gamma _I}
$
is a
partition
of
the estimated
observed information matrix,~$\mathbf{I}(\widehat{\bm{\gamma}})$,
limited to the
estimates of interest.
Under~$\mathcal{H}_0$,
the test statistic,~$T_W$,
has asymptotically
chi-squared distribution
with~$\nu$
degrees of freedom,
$\chi^2_\nu$.
Thus,
the proposed detector
consists of
comparing the computed value
of~$T_W$
with
a threshold value~$\varepsilon$.
The
threshold value
is
obtained
from
the~$\chi^2_\nu$ distribution
and
the
desired
probability of false alarm~\cite{Kay1998-2}.

To
illustrate
the above
approach,
we consider
the problem of detecting
a signal~$s[n]$
embedded
in noise
from a
measured signal~$y[n]$.
For such, we have
the following
BBARMA model:
\begin{align}
\begin{split}
\label{e:modeldec}
g(\mu[n]) =
\zeta +
s [n] \beta_1
+ \sum \limits _{i=1} ^p \phi_i y^\star [n-i]
+ \sum \limits _{j=1} ^q \theta_j \{ y^\star [n-j] - \mu[n-j] \}
,
\end{split}
\end{align}
where~$\beta_1$
is the unknown
amplitude of the signal~\cite{Kay1998-2}.
To detect
whether~$s[n]$
is present,
we have
the following hypotheses:
\begin{align}
\label{e:hip}
\begin{cases}
\mathcal{H}_0 : \beta_1=0
,
\\
\mathcal{H}_1 : \beta_1 \neq 0
.
\end{cases}
\end{align}
The detector is derived using
the
Wald test described above.
We
reject~$\mathcal{H}_0$
when~$T_W > \varepsilon$~\cite{Kay1998-2}.
In this situation,~$\beta_1 \neq 0$,
indicating the presence of the signal.

\section{Tools and Diagnostic Analysis }
\label{s:forecasting}

The data forecasting
for the
BBARMA$(p,q)$
model
can be derived by applying
the
CMLE of
${\bm{\gamma}}$,~$\widehat{\bm{\gamma}}$,
to obtain estimates~$\widehat{\mu}[n]$,
for~$\mu[n]$.
The mean response estimate at~$N+h$,
with~$h=1,2,\ldots, H$,
where~$H$ is the
forecast horizon,
is given by
\begin{align*}
\widehat{\mu}[N+h]
&=
g^{-1} \left(  \widehat{\zeta}
+\mathbf{x}^\top[n] \widehat{\bm{\beta}} +
\sum\limits_{i=1}^{p}\widehat{\phi}_i \lbrace y ^\ast [N+h-i]  \rbrace
+
\sum\limits_{j=1}^{q}\widehat{\theta}_j\lbrace r^\ast[N+h-j] \rbrace \right)
,
\end{align*}
where
\begin{align*}
y^\ast[N+h-i]  & =\left\{\begin{array}{rc}
\widehat{\mu}[N+h-i],&\textrm{if}\;\;\; i< h,\\
y^\star[N+h-i], &\textrm{if}\;\;\; i\geq h,
\end{array}\right.\\
r^\ast[N+h-j] &=\left\{\begin{array}{rc}
0,&\textrm{if}\;\;\; j< h,\\
\widehat{r}[N+h-j], &\textrm{if}\;\;\; j\geq h,
\end{array}\right.
\end{align*}
and $\widehat{r}[n] = y^\star [n] - \widehat{\mu}[n]$.
The quantity~$\widehat{\mu}[n] \in (0,1)$
can be mapped
to the discrete
set~$\left\lbrace 0,1,2,\ldots,K \right\rbrace$
by means
of~$\text{round}(\widehat{\mu}[n] \cdot K)$,
where~$\text{round}( \cdot )$ is a round function.
Note that
no parameter restrictions
are required
for fitting
or forecasting
based on BBARMA model.

The correct adjustment of
the proposed model
is important
to obtain
accurate
out-of-signal forecasting.
For the BBARMA model selections,
we adopted
the following
information criteria:
Akaike's~(AIC)~\cite{Akaike},
Schwartz's~(SIC)~\cite{schwarz1978},
and Hannan and Quinn's~(HQ)~\cite{hannan1979}.
Residuals
are useful for
performing the
diagnostic analysis of the fitted model
and
can be defined as
a function of the
observed
and
predicted values
of the model~\cite{kedem2005,brockwell2013}.

Different
types of residuals are considered in literature
for several classes of
models,
such as
ordinary residuals,
standardized residuals
and
some
residuals
in the predictor scale.
We employed
the
standardized ordinary
residuals
\begin{align*}
\epsilon[n]
& =
\frac{y^\star [n] - \widehat{\mu}[n]}
{\sqrt{\widehat{\operatorname{Var}}(y[n])}}
,
\end{align*}
where
$\widehat{\operatorname{Var}}(y[n])
=
K \widehat{\mu}[n](1-\widehat{\mu}[n])
\left[ \frac {K + \widehat{ \varphi}}
{1+\widehat{\varphi}} \right]$.

A good
model adjustment
is indicated by
zero mean and constant variance
of
the standardized residual~\cite{kedem2005}.
Also, it is expected
that the
autocorrelation and partial autocorrelation
and conditional heteroscedasticity
in the series of residuals
are absent~\cite{Box2008}.
The residual
autocorrelation function~(ACF)
is given by
\begin{align*}
\widehat{\rho} _k
=
\frac{\sum \limits_{n=m+1}^{N-k}(\epsilon[n] - \bar{\epsilon})
	(\epsilon[n+k] - \bar{\epsilon})}
{\sum \limits_{n=m+1}^{N-k}(\epsilon[n] - \bar{\epsilon})^2}
,
\quad
k = 0,1,\ldots
,
\end{align*}
where~$\bar{\epsilon} = (N-m)^{-1} \sum _{n=m+1}^N \epsilon[n]$.
The distribution of~$\widehat{\rho} _k $
is approximately
normal with zero mean and variance~$1 / (N-m)$,
for~$i > 1$ and~$N \rightarrow \infty $~\cite{anderson1942,kedem2005,Box2008}.
Lagrange Multiplier~\cite{Engle1982},
Box-Pierce~\cite{BoxandPierce1970},
Ljung-Box~\cite{LjungandBox1978},
and the
ACF plot
are useful
to
verify whether the
autocorrelation and the conditional
heteroscedasticity
in the series of residuals
are absent.

\section{Numerical Results}
\label{s:vali}

In this section,
we
aim at
(i)~evaluating the CMLE of
the parameters of
the BBARMA model
and
(ii)~assessing
the performance
of the proposed detector
and the BBARMA model prediction.
For such,
computational
experiments
based on
Monte Carlo simulations
were considered.
Additionally,
the derived BBARMA model
was employed to predict
the
monthly number of rainy days
in Recife, Brazil.

\subsection{Evaluation of the CMLE}

Signals $y[n]$
were generated
from the beta binomial distribution
by the
acceptance-rejection method~\cite{Casella2004}
with mean given by~\eqref{model},
logit link function, $K = 255$
(8-bit signals),
without covariates in the
simulations.
We
considered
simulations
under
two scenarios.
Scenario~I
employs~$\mu \approx 0.9$
(asymmetric distribution)
and
Scenario~II
adopts~$\mu \approx 0.5$
(almost symmetric distribution).
For such,
the
selected
parameters
were~$\zeta= 1$,~$\phi _1 = 1$,~$\theta=0$,
and~$\varphi = 20$,
for
Scenario~I,
and~$\zeta= 0.2$,~$\phi _1 = 0.5$,~$\theta _1 = 0.3$,
and~$\varphi = 15$,
for
Scenario~II.
The number of Monte Carlo replications
was set to~$10,000$ and the
signal
lengths
considered
were~$N \in \{
150, 300, 500
\}$.
In order to numerically evaluate the point estimators,
we computed
the
mean,
bias,
and
mean square error~(MSE)
of the CMLE.

For the evaluation of interval estimation,
we calculated
the
coverage rates~(CR)
of the
confidence interval~(CI)
with confidence~$100(1-\alpha)\%$.
The CR
is derived based
on the
CMLE asymptotic distribution
and
it
is
defined as
\begin{align*}
\left[\widehat{\gamma}_i - z_{1-\alpha/2}
\sqrt{\mathbf{I}^{-1}_{ii}(\widehat{\bm{\gamma}})};
\widehat{\gamma}_i + z_{1-\alpha/2}
\sqrt{\mathbf{I}^{-1}_{ii}(\widehat{\bm{\gamma}})}\right],
\end{align*}
where~$\gamma_i$,
$i = 1 , 2, \ldots , l+p+q+2$,
denotes the~$i$th component of~$\bm{\gamma}$,
$\mathbf{I}_{ii}^{-1}(\widehat{\bm{\gamma}})$
is the~$i$th element of the
diagonal of~$\mathbf{I} ^{-1}(\widehat{\bm{\gamma}})$,~$\alpha$
is the significance level,
and~$z_{\varrho}$
is
the~$\varrho$th quantile of the
standard normal distribution.
For each Monte Carlo replication,
we computed the~CI
and
interrogated whether the~CI contains the true parameter
or not.
The~CR
is given by the percentage
of replications
for which
the
parameter
is
in the~CI.

Tables~\ref{t:bbar} and~\ref{t:bbarma}
present the simulation results
for Scenarios~I and~II,
respectively.
As expected,
both
bias and MSE
figures
improved
as~$N$
grows.
This behavior
agrees with the
asymptotic property
(consistency)
of the CMLE.
The
CR values
of the
BBARMA$(1,0)$ model
are close to the nominal value
of~$0.90$,
for all considered~$N$.
In accordance with the literature~\cite{Ansley1980,Palm2017},
the BBARMA$(1,1)$ model
presents CR values
close
to~$0.90$
for larger signal lengths.
Convergence failures were absent
for all the tested scenarios.

\begin{table}
\caption{
Simulation results on point
and
interval estimation
of the~BBARMA$(1,0)$ model,
considering
a significance level for~$\alpha = 10\%$
-- Scenario~I
}
\label{t:bbar}
\centering
\begin{tabular}{lccc}
\toprule
Measures 	&	$\widehat{\zeta}$ 	&	$\widehat{\phi}_1$ &
$\widehat{\varphi}$ 	\\
\midrule
\multicolumn{4}{c}{ $N=150$ } \\
\midrule
Mean &  $1.0852$ & $0.9036$ &  $20.6377$  \\
Bias &  $0.0852$ & $-0.0964$ &  $0.6377$ \\
MSE   & $0.3091$ &  $0.4104$  &  $7.8154$   \\
CR &   $0.9045$ &  $0.9045$ & $0.9015$  \\
\midrule
\multicolumn{4}{c}{ $N=300$ } \\
\midrule
Mean &  $1.0451$ & $0.9493$ & $20.3392$  \\
Bias &  $0.0451$  &  $-0.0507$  & $0.3392$ \\
MSE   & $0.1538$ &  $0.2046$ &  $3.5039$ \\
CR & $0.9011$ &  $0.9013$ &  $0.9039$ \\
\midrule
\multicolumn{4}{c}{ $N=500$ } \\
\midrule
Mean & $1.0344$ & $0.9611$ & $20.2202$  \\
Bias &  $0.0344$ &  $-0.0389$ & $0.2202$ \\
MSE   & $0.0929$ & $0.1235$ &  $2.0854$ \\
CR & $0.8980$ & $0.8964$ & $0.8996$  \\
\bottomrule
\end{tabular}
\end{table}

\begin{table}
\caption{
Simulation results on point
and
interval estimation
of the~BBARMA$(1,1)$ model,
considering
a significance level for~$\alpha = 10\%$
-- Scenario II
}
\label{t:bbarma}
\centering
\begin{tabular}{lcccc}
\toprule
Measures 	&	$\widehat{\zeta}$ 	&	$\widehat{\phi}_1$ 	&
$\widehat{\theta}_1$ &$\widehat{\varphi}$ 	\\
\midrule
\multicolumn{5}{c}{ $N=150$ } \\
\midrule
Mean &  $0.2987$  &  $0.3428$ &  $0.4614$ & $15.5923$ \\
Bias & $0.0987$ &  $-0.1572$ & $0.1614$ & $0.5923$ \\
MSE   & $0.9607$ &  $2.4550$ & $2.5562$ &  $4.1275$ \\
CR & $0.7601$ &  $0.7593$ &  $0.7523$ &  $0.8949$ \\
\midrule
\multicolumn{5}{c}{ $N=300$ } \\
\midrule
Mean & $0.2629$ & $0.4007$ &  $0.3971$ &  $15.2820$ \\
Bias & $0.0629$ &  $-0.0993$ & $0.0971$ &  $0.2820$ \\
MSE   & $0.6209$ &  $1.5866$ &  $1.6179$ & $1.7988$ \\
CR & $0.8105$ &  $0.8085$ &  $0.8064$ &  $0.8998$ \\
\midrule
\multicolumn{5}{c}{ $N=500$ } \\
\midrule
Mean &  $0.2393$ &   $0.4375$ &  $0.3602$ &  $15.1728$  \\
Bias & $0.0393$ & $-0.0625$ &  $0.0602$ & $0.1728$ \\
MSE   &  $0.4106$  &  $1.0514$ &  $1.0704$ &  $1.0238$ \\
CR &  $0.8405$ &  $0.8405$ &  $0.8367$ &  $0.9031$  \\
\bottomrule
\end{tabular}
\end{table}

\subsection{Evaluation of the Proposed Detector}

Considering the same
simulations parameters
as detailed in the
previous subsection,
we aim at assessing
the performance
of the introduced
detector showed in~\eqref{e:hip}.
For such,
in~\eqref{e:modeldec}, we adopted~$p=q=1$.
The interest signal,~$s[n]$,
was selected
as~$s[n] = \cos ( 2 \pi f_0 n) $,
where~$f_0$
is the signal frequency.
This is the classical
sinusoidal detection problem
which is present
in many fields,
such
as
radar,
sonar,
communication
systems~\cite{Kay1998-2},
and
discrete-time signal
seasonality detection~\cite{bloomfield2004fourier}.

\begin{figure}
\centering
\subfigure[Simulated digital signal -
Scenario III]{
	\includegraphics[scale=0.45]{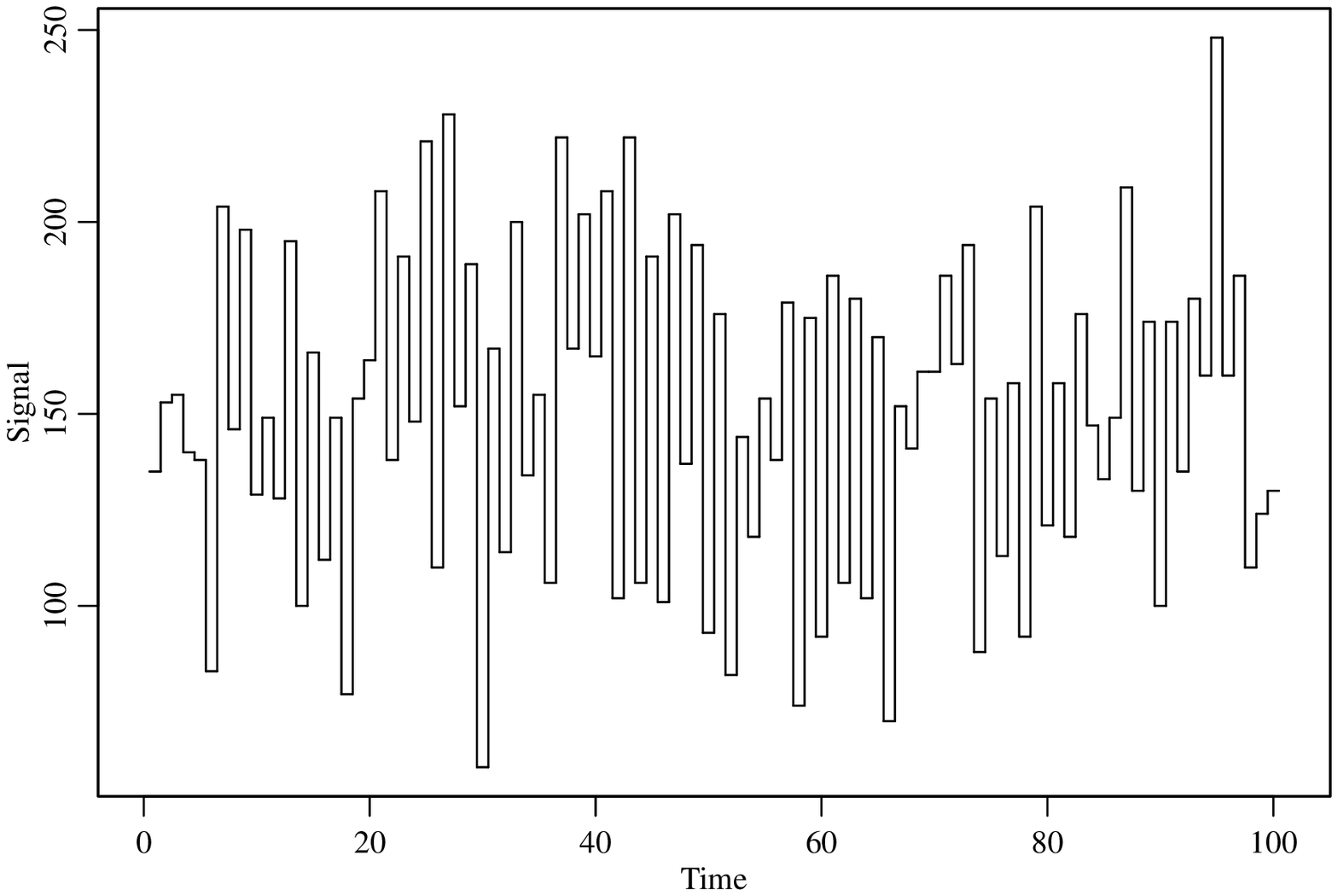}}
\subfigure[Simulated digital signal -
Scenario IV]{
\includegraphics[scale=0.45]{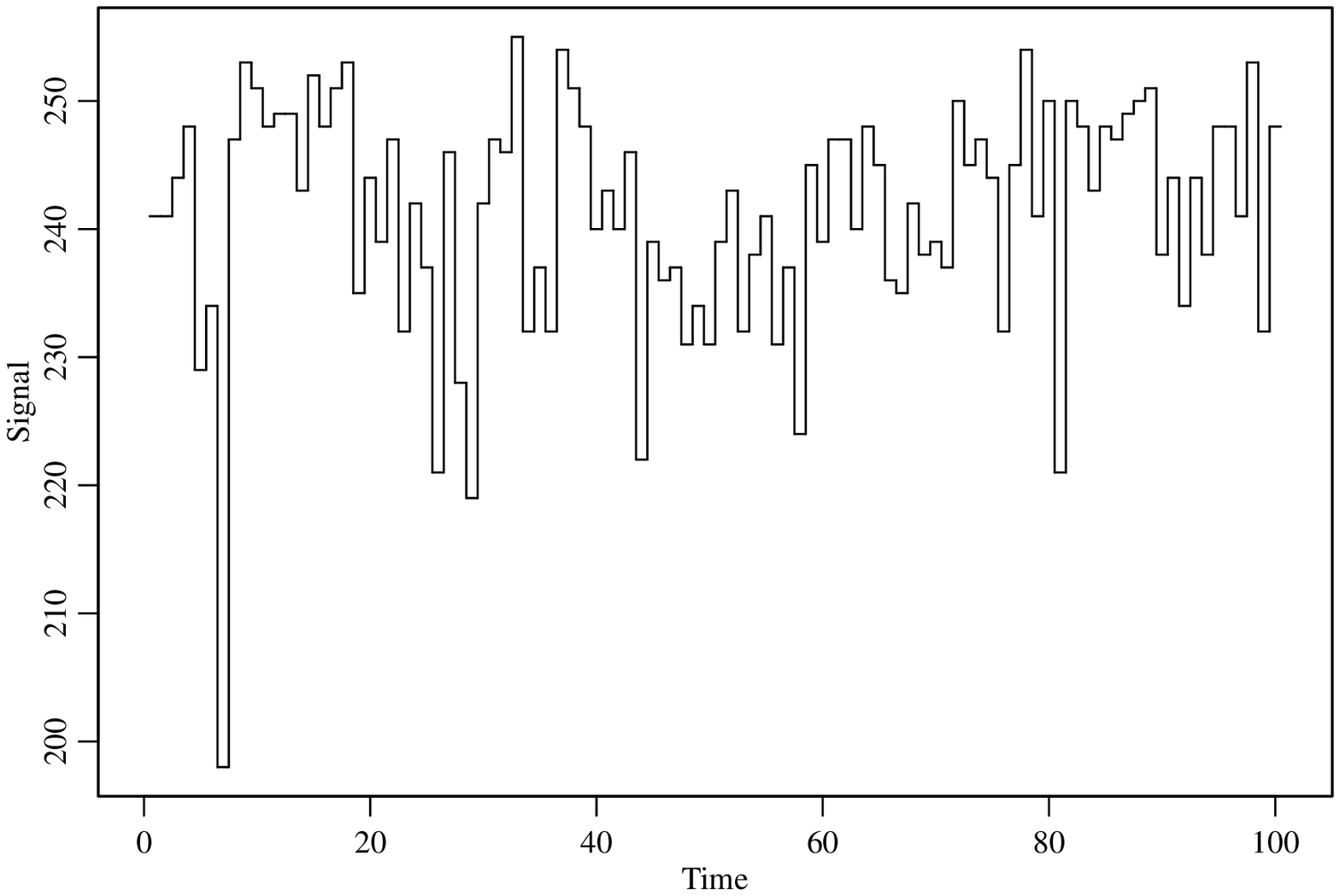}}
\caption{Simulated digital signals
tested in the proposed detector,
considering
Scenarios~III and~IV.
}\label{f:sim}
\end{figure}

For each Monte Carlo replication,
we fitted  the BBARMA model
in a simulated digital signal~$y[n]$ as follows
\begin{align*}
g(\mu[n]) =
\zeta +
s [n] \beta_1
+ \phi_1  y^\star [n-1]
+  \theta_1\{ y^\star [n-1] - \mu[n-1] \}
.
\end{align*}
We
considered
two scenarios
to obtain~$y[n]$:
(i)~Scenario~III
employed~$\zeta= 0.2$,~$\beta_1=0.5$,~$\phi _1 = 0.5$,~$\theta _1 = 0.3$,~$\varphi
= 15$,
and~$f_0=0.5$;
and
(ii)~Scenario~IV
set~$\zeta= 1$,~$\beta_1= 0.1$,~$\phi _1 = 2$,~$\theta _1 =1$,~$\varphi =50 $,
and~$f_0= 0.7$.
Scenarios~I and~II employed~$\mu \approx 0.5$
and~$\mu \approx 0.9$, respectively.
The
levels of significance
were~$\alpha \in \{0, 0.05, 0.1, 0.15, 0.2, 0.3 ,
0.4, 0.5, 0.6, 0.7, 0.8, 0.9, 1 \}$,
the number
of Monte Carlo replication
equal to~$5,000$,
and~$N=100$.
Figure~\ref{f:sim}
shows
a typical
realization
of the simulated signals,
considering
Scenarios~III and~IV.

Based on~\eqref{e:hip},
a signal is detected
when
the amplitude~$\beta_1 \neq 0$,
i.e.,
when~$\widehat{\beta}_1$
is significant to the model
and the null hypothesis
in~\eqref{e:hip} is rejected.
To estimate the probability of
detection,
we computed the proportion
of Monte Carlo replications
in which~$\widehat{\beta}_1$
was
significant to the model.
We used
the empirical size
of the hypothesis test,
which is
computed
according to
the following
algorithm:
(i)~for each Monte Carlo replication,
generate a signal,~$y^{\ast}[n]$,
without covariates;
(ii)~fit the BBARMA model
for~$y^{\ast}[n]$,
including the parameter~$\beta_1$;
and
(ii)~compute
the percentage
of replications that~$\beta_1 \neq 0$.

\begin{figure}
\centering
\subfigure[Scenario~I]{
\includegraphics[scale=0.44]{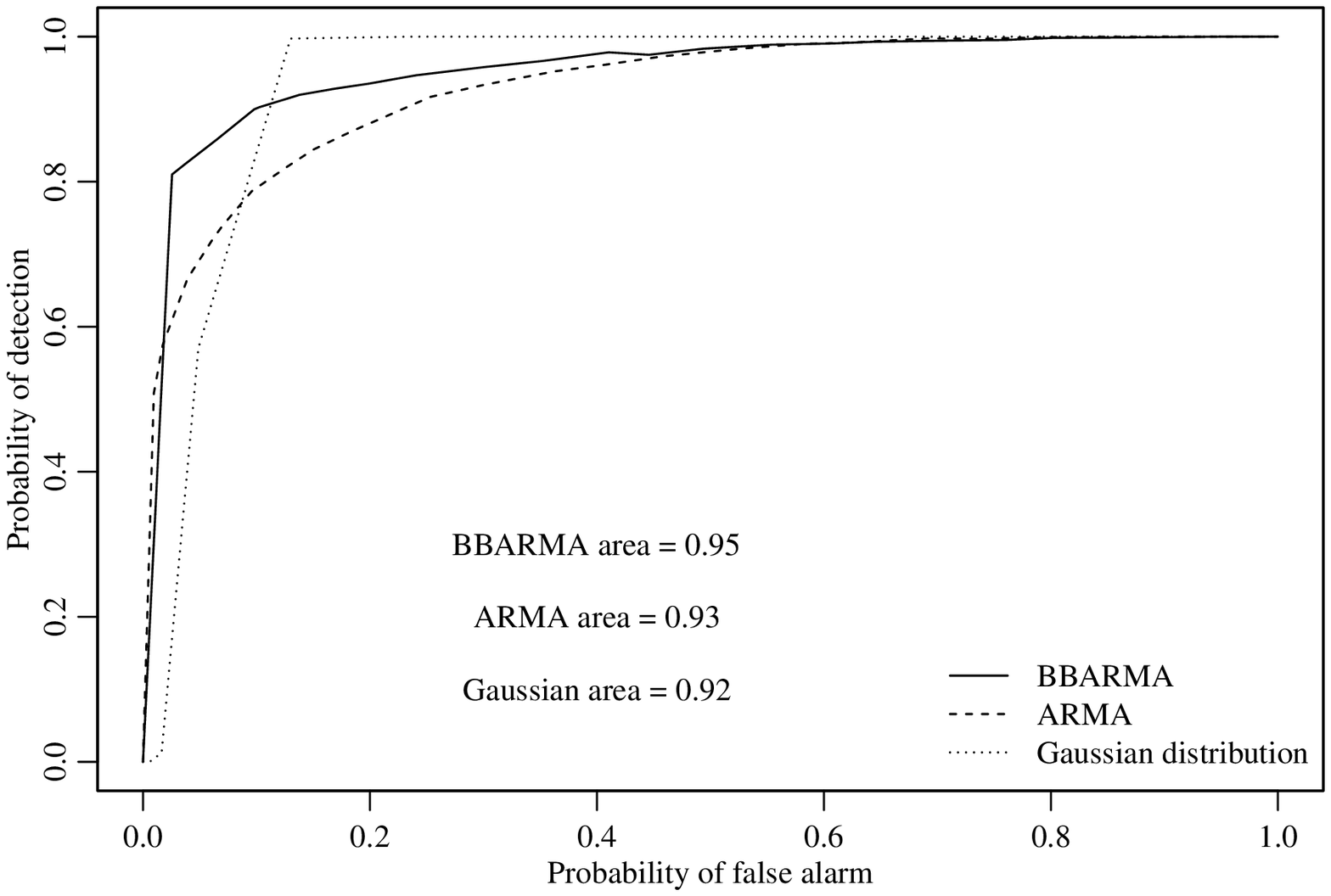}
\label{f:roc1}}
\subfigure[Scenario~II]{
	\includegraphics[scale=0.44]{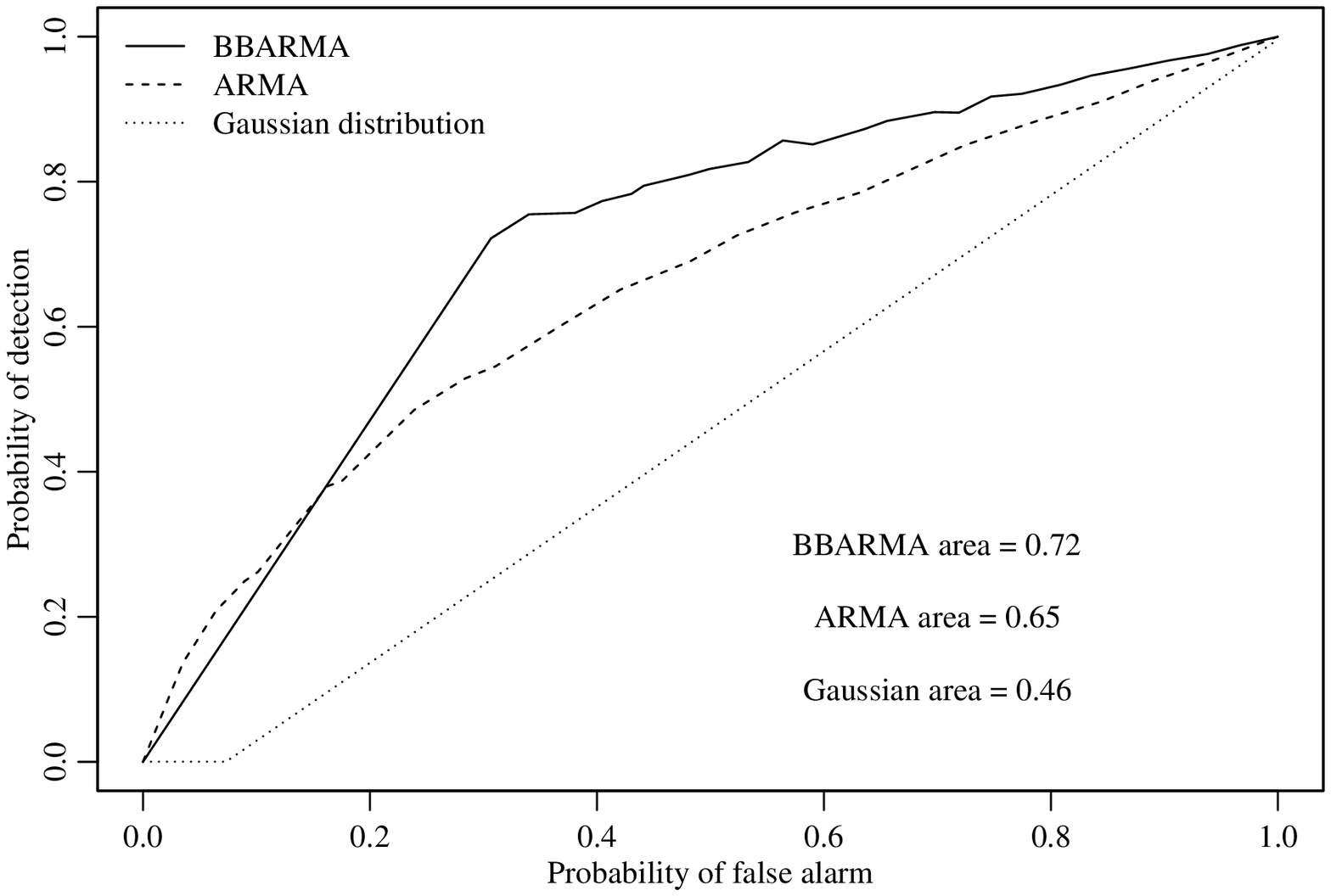}
	\label{f:roc2}}
\caption{ROCs of the detection results for Scenarios~I and~II,
comparing the BBARMA-, ARMA-, and Gaussian-based detectors.
}
\end{figure}

We compared the proposed detector
with the
widely popular
ARMA-
and
Gaussian-based detectors~\cite{Kay1998-2}.
Figures~\ref{f:roc1}
and~\ref{f:roc2} present
the
receiver operating characteristic~(ROC)
curves~\cite{ROC}
of the detection results,
showing the probability of detection
versus the estimated probability of false alarm.
The proposed BBARMA detector
outperformed
the competing methods
in terms of
probability of detection
and estimated probability of false alarm
in both considered scenarios.
The area under the ROC curves
for the ARMA detector
was~$2.10\%$
and~$9.72\%$
lower when compared
with ROC curve of the proposed detector
under
Scenario~III and~IV,
respectively;
for the Gaussian detector,
the ROC area values were
also smaller,
being~$3.16\%$
and~$36.11\%$
lower,
for
Scenario~III and~IV,
respectively.

\subsection{Monthly Number of Rainy Days Prediction }
\label{s:aplicacao}

In this section,
we submit actual measured data to the proposed method.
We applied the
BBARMA model
to monitor
the number of rainy days per month
in Recife, Brazil,
since this topic
is
gaining importance
in discrete-time
signal modeling~\cite{gouveia2018,tian2016,lacombe2014,ehelepola2015}.
Additionally,
we performed a
seasonal detection
due its importance
in the context
of discrete-time signal analysis,
as discussed, for example,
in~\cite{verbesselt2010detecting,
	hirsch1984nonparametric,bloomfield2004fourier}.
The employed signal values
are available in~\cite{data}.
The
data were
measured
in the period of
February 2006 to
March 2013,
consisting of
86~monthly observations.
The ending~$12$ and~$15$
observations were separated
for
assessment of the BBARMA model forecasting performance.

\begin{figure}
	\centering
	\subfigure[Observed data]
	{\label{f:serie2}\includegraphics[width=0.45\textwidth]{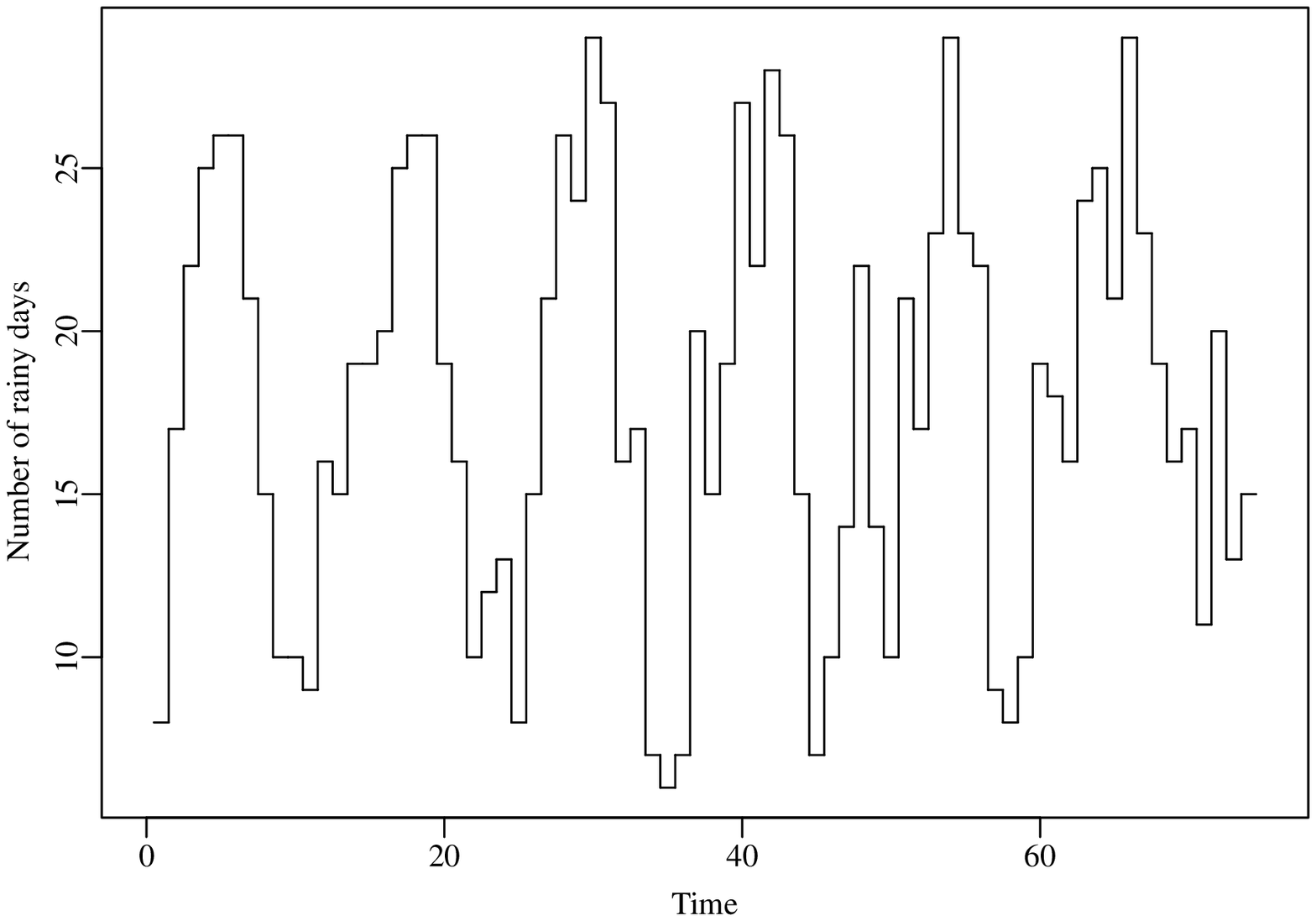}}
	\subfigure[Seasonality]
	{\label{f:sazo}\includegraphics[width=0.45\textwidth]{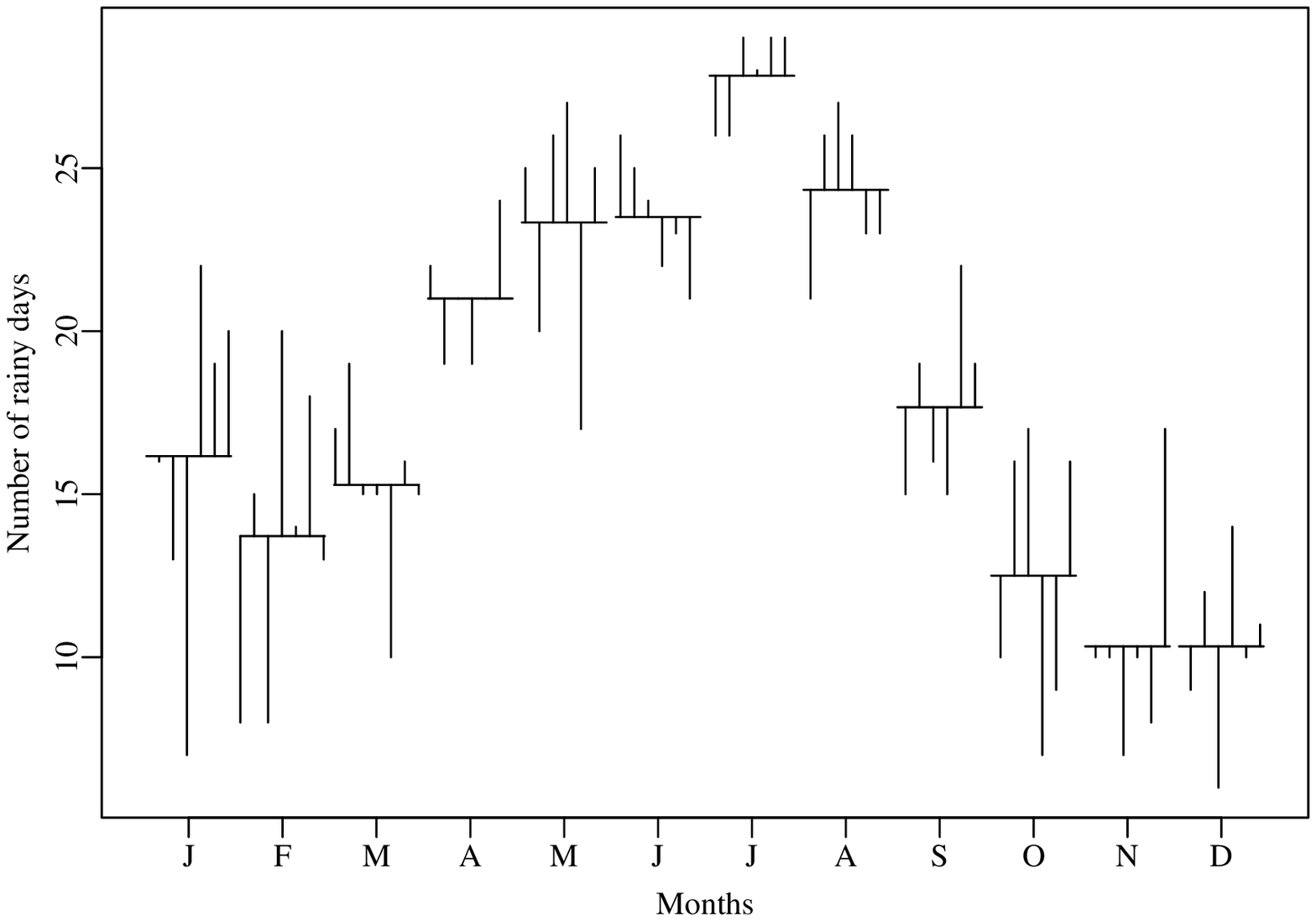}}
	\subfigure[Sampling ACF]
	{\label{f:fac2}\includegraphics[width=0.45\textwidth]{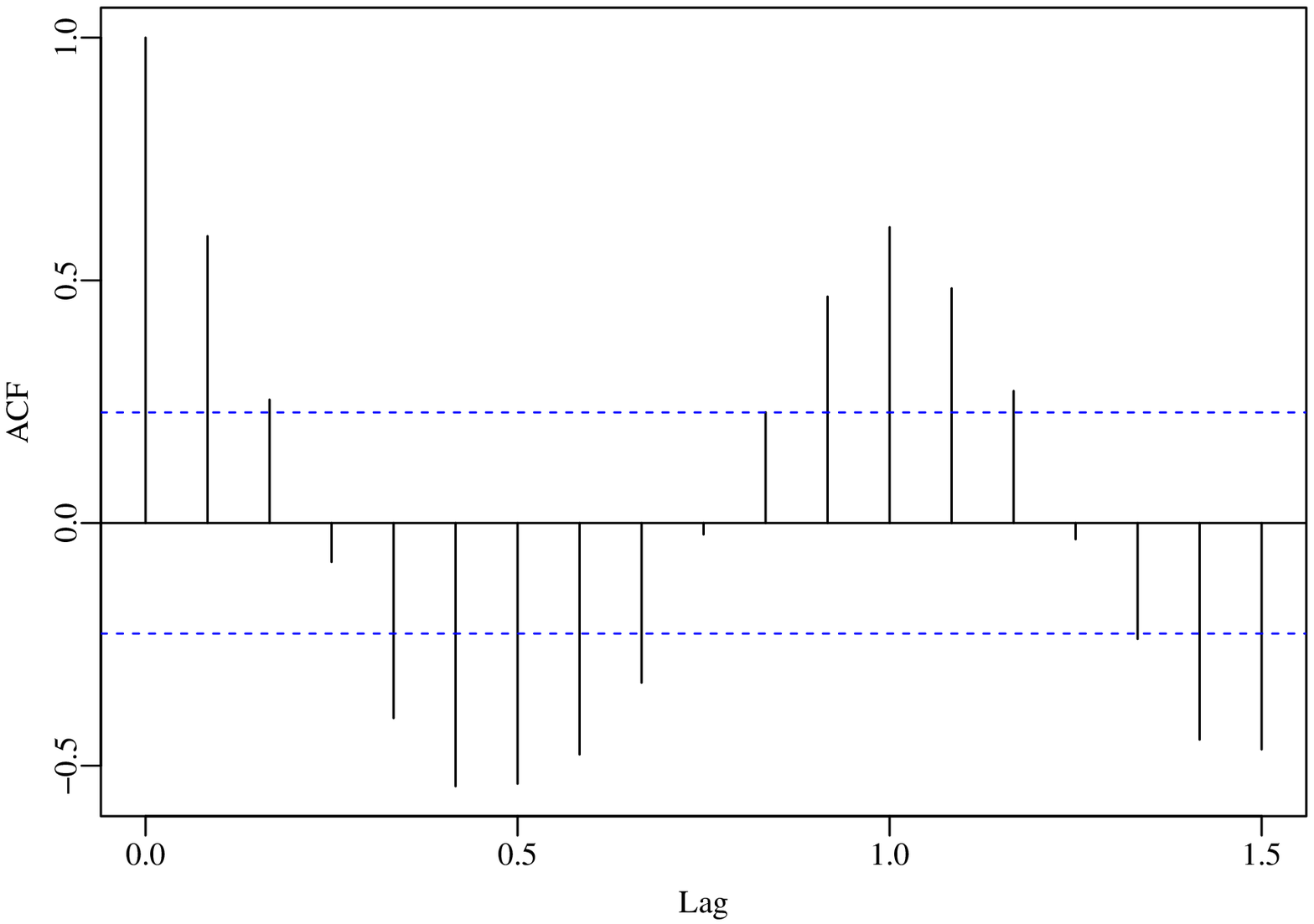}}
	\subfigure[Sampling PACF]
	{\label{f:facp2}\includegraphics[width=0.45\textwidth]{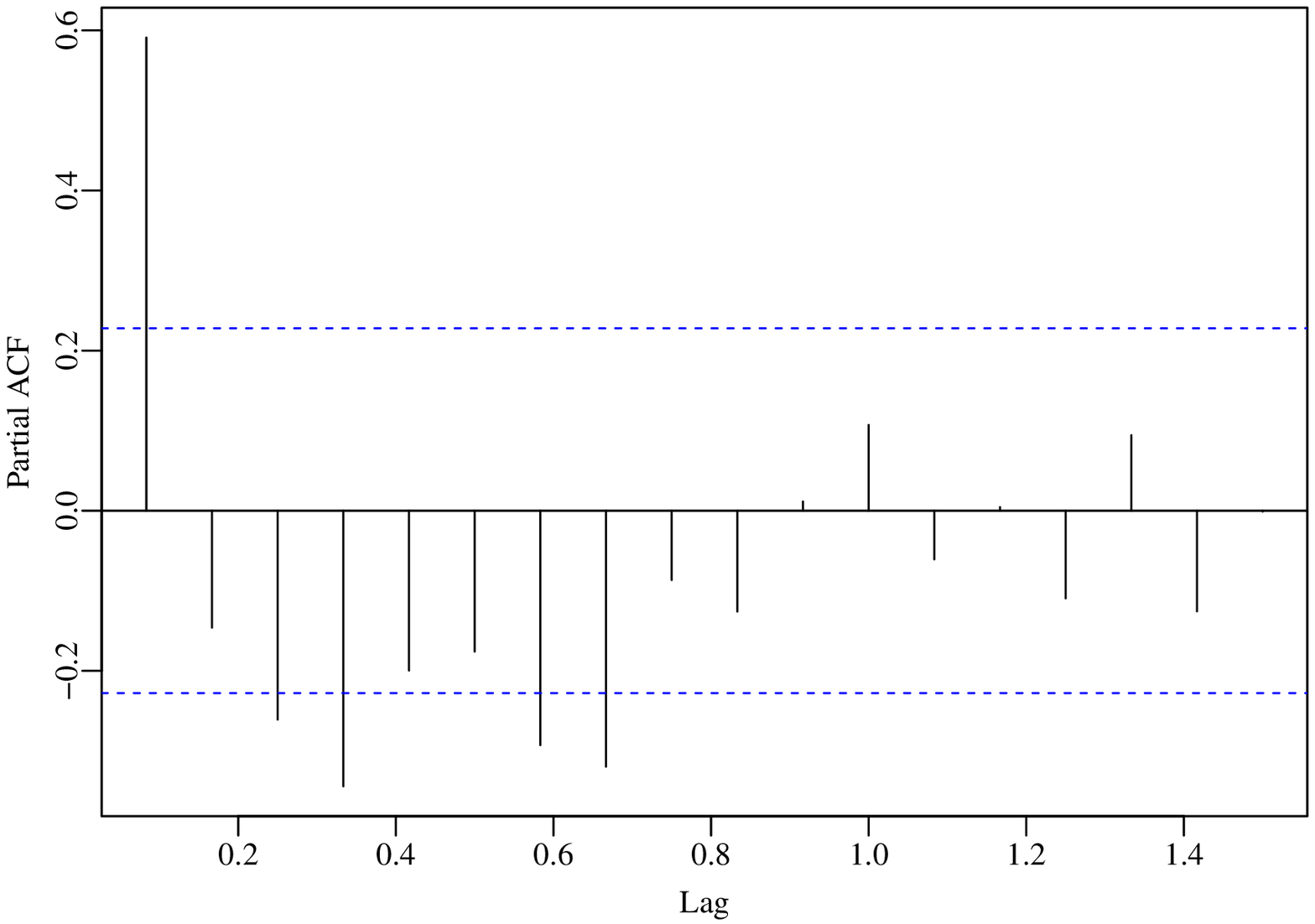}}
	\caption{
		Line chart, seasonal component,
		and correlograms of the
		discrete-time signal
		of the monthly number of rainy days
		in Recife, Brazil.}
	\label{f:dados2}
\end{figure}

Figures~\ref{f:serie2}
and~\ref{f:sazo}
display
the considered
signal
with
unconditional mean
of~$17.90$
and
the
seasonal component in the data,
respectively.
The maximum value
of the
signal
was~$29$,
observed in July 2008,
and the lower precipitation amount
was~$6$, in December 2008.
Figures~\ref{f:fac2} and~\ref{f:facp2}
present
the sampling ACF
and the sampling partial autocorrelation function~(PACF),
respectively.
To account the monthly seasonal component
showed in
Figure~\ref{f:sazo},
we
introduced the~$\cos(2 \pi n / 12)$
covariate,
for~$n \in \lbrace 1, 2, \ldots , N \rbrace$,
in the
harmonic regression approach
suggested by~\cite{bloomfield2004fourier}.
Consequently,
the proposed detector
can also be used
to
verify the presence
of the seasonal component
in the employed discrete-time signal.
Based on~\eqref{e:hip},
the seasonal component is detected
when~$\beta_1 \neq 0$,
i.e.,
when the estimated parameter~$\widehat{\beta}_1$
associated to the
sinusoidal
covariate
is significant to the model
and the null hypothesis
in~\eqref{e:hip} is rejected.

Based on
the three-stage iterative Box-Jenkins methodology~\cite{Box2008},
i.e.,
identification---considering
an exhaustive search
aiming at
minimizing the AIC---estimation,
and
diagnostic checking,
we
successfully modeled
the employed data using the BBARMA$(0,1)$
model with the sinusoidal covariate
given above,
and
considering the logit link function.
We employed comparisons among
the forecast values
of the proposed BBARMA model,
the ARMA model~\cite{Box2008},
and
the Holt-Winters method~\cite{holt2004forecasting,winters1960forecasting}.
The ARMA model
is widely used in time series analysis
and
the Holt-Winters method
is a nonparametric  approach
for discrete-time signal processing.
Following the
iterative methodology described above,
the
ARMA$(2,1)$ model
and the Holt-Winters additive method
were elected to fit the considered data set.

The diagnostic analysis of the
fitted
model
was
based on the standardized residual~$\epsilon[n]$.
Table~\ref{T:ajuste12} presents
the fit of the selected model and the diagnostic analysis.
The $p$-values of the Wald test
presented in Table~\ref{T:ajuste12}
suggest that the estimated parameter~$\widehat{\beta}_1$
is significant to the model for a probability of
false alarm equal to~$0.05$.
Hence, the null hypothesis in~\eqref{e:hip}
can be rejected, indicating a correct detection
of the season component.
Additionally,
the residuals of the fitted model do not exhibit
conditional heteroscedasticity or autocorrelation,
indicating that the fitted model can be safely used for
out-of-signal forecasting.

\begin{table}
	\centering
	\tablesize
	\caption{
		BBARMA$(0,1$)~model adjusted for the monthly number of rainy days
		in Recife, Brazil,
considering~$H=12$}
	\label{T:ajuste12}
	\begin{tabular}{lcccc}
		\toprule
		&${\zeta}$ & ${\beta}_1$ &  ${\theta}_1$ &
		${\varphi}$\\
		\midrule
		Estimator &   $0.3555$ &   $-1.0427$ &    $0.7470$ &   $27.3863$        \\
		Standard Error &  $0.0572$ &    $0.0815$ &    $0.4130$ &    $8.5568$    \\
		$p$-value &  $<0.001$ &   $<0.001$ &   $0.0705$ &    $0.0014$ \\
		\midrule
		\multicolumn{5}{c}{Diagnostic analysis}\\
		\midrule
		\multicolumn{2}{c}{Test} &  &&$p$-value \\
		\midrule
		\multicolumn{2}{c}{Lagrange Multiplier}   && & $0.3582$ \\
		\multicolumn{2}{c}{Box-Pierce} & & &  $0.1881$ \\
		\multicolumn{2}{c}{Ljung-Box}   & && $0.1295$ \\
		\bottomrule
	\end{tabular}
\end{table}

\begin{figure}
	\begin{center}
		\subfigure[Observed and predicted values]
		{\label{f:adjusted}\includegraphics[width=0.45\textwidth] {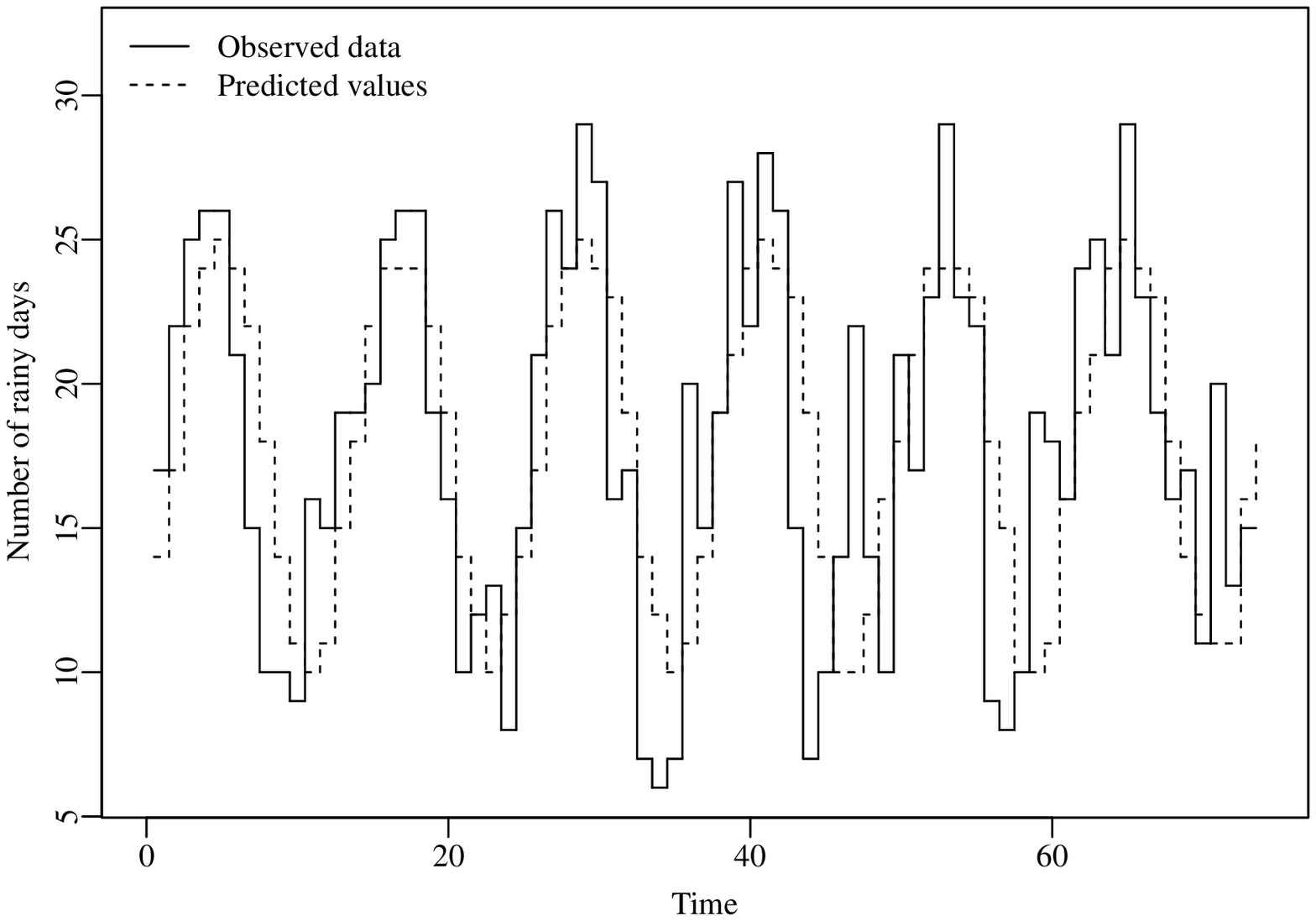}}
		\subfigure[Out-of-signal forecast]
		{\label{f:prev}
			\includegraphics[width=0.45\textwidth] {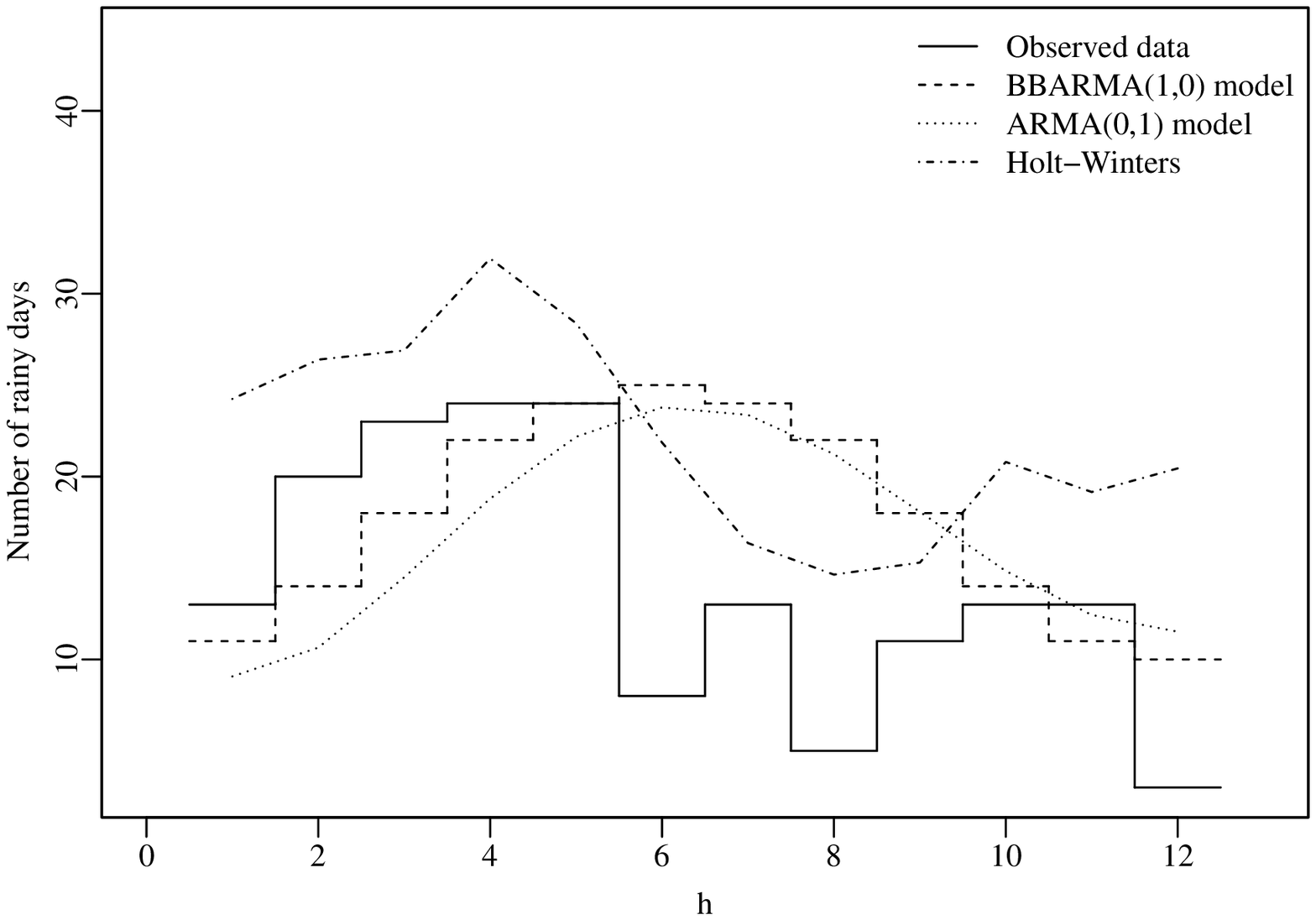}}
		\caption{
			Fitted values and out-of-signal forecast of the
			fiited BBARMA$(1,0)$ model
			for~$H=12$.}
	\end{center}
\end{figure}

Figure~\ref{f:adjusted}
shows the observed and predicted values
obtained considering the fitted BBARMA model.
For instance,
Figure~\ref{f:prev}
presents
the out-of-signal forecast
of the adjusted BBARMA$(0,1)$ model,
ARMA$(2,1)$ model,
and Holt-Winters method
for~$H=12$.
The observed data
and the BBARMA model forecasting
are
discrete values and are represented
by
line segments between pairs of points.
The ARMA model and Holt-Winters
out-of-signal forecasts
are real numbers
whose values
are
joined
with lines segments.
In order to have a
meaningful
comparison
of the predicted values,
we computed some  goodness-of-fit measures,
such as
root mean
squared error (RMSE),
median absolute error~(MdAE),
and
mean absolute scaled error~(MASE);
such measures are expected
to be as close to zero as possible~\cite{Hyndman2006}.
As shown in Table~\ref{t:comp},
the proposed
model outperforms the ARMA model
and Holt-Winters method according to the considered figures of merit
in both evaluated forecast horizon.

\begin{table}
	\caption{
		Forecasting performance comparison among the BBARMA model,
		ARMA model, and Holt-Winters method.
Best results are highlighted
}
	\label{t:comp}
	\tablesize
	\begin{center}
		\begin{tabular}{lccc}
			\toprule
	& RMSE & MdAE & MASE \\
			\midrule
	\multicolumn{4}{c}{$H=12$} \\
			\midrule
		BBARMA & $\textbf{8.5196}$ &  $\textbf{5.5000}$ &  $\textbf{1.2170}$ \\
			ARMA & $8.9069$ & $7.7899$ &  $1.4098$ \\
			Holt-Winters & $9.0478$ &  $7.0965$ &  $1.5235$  \\
						\midrule
			\multicolumn{4}{c}{$H=15$} \\
			\midrule
			BBARMA & $\textbf{6.1319}$ & $\textbf{2.0000}$ &    $0.8928$ \\
			ARMA &  $6.1443$ & $2.0945$ & $\textbf{0.8747}$\\
			Holt-Winters & $8.5654$ &  $6.2937$ &  $1.4856$   \\
			\bottomrule
		\end{tabular}
	\end{center}
\end{table}

\section{Conclusion}
\label{s:conclusion}

In this paper,
we derived the BBARMA model
and a signal detector
based on the asymptotic properties
of the discussed
model estimators.
We
introduced
an inference approach
for
the model parameters,
diagnostic measures,
out-of-signal forecasting,
the
conditional
observed information
matrix,
and
the
asymptotic properties of the CMLE.
Monte Carlo simulations
were used as a tool
to
evaluate the performance
of the CMLE and of the proposed
signal detector,
indicating the consistency of the CMLE.
The proposed BBARMA detector
could
outperform the
ARMA-
and
Gaussian-based detectors
in the evaluated scenarios.
The proposed model is presented
as a suitable tool
for quantized signal detection.
Additionally,
an experiment of the derived BBARMA model
considering the
monthly number of rainy days
in Recife, Brazil, is presented and discussed.
The BBARMA model
detected the component seasonal
in the data set
and
presented
more accurate forecasting results
than the traditional ARMA model and Holt-Winters method.

\section*{Acknowledgements}

We gratefully
acknowledge partial financial support from
Conselho Nacional de Desenvolvimento Cient\'ifico and Tecnol\'ogico (CNPq),
and Coordena\c{c}\~ao de Aperfei\c{c}oamento
de Pessoal de N\'ivel Superior (CAPES), Brazil.

{\small
\singlespacing
\bibliographystyle{siam}
\bibliography{betareg}
}

\end{document}